# Chapter 1

# Stochastic spatial Lotka–Volterra predator-prey models

Uwe C. Täuber

*Department of Physics & Center for Soft Matter and Biological Physics,*
*850 West Campus Drive (MC 0435), Faculty of Health Sciences,*
*Virginia Tech, Blacksburg, VA 24061, USA, tauber@vt.edu*

Dynamical models of interacting populations are of fundamental interest for spontaneous pattern formation and other noise-induced phenomena in nonequilibrium statistical physics. Theoretical physics in turn provides a quantitative toolbox for paradigmatic models employed in (bio-)chemistry, biology, ecology, epidemiology, and even sociology. Stochastic, spatially extended models for predator-prey interaction display spatio-temporal structures that are not captured by the Lotka–Volterra mean-field rate equations. These spreading activity fronts reflect persistent correlations between predators and prey that can be analyzed through field-theoretic methods. Introducing local restrictions on the prey population induces a predator extinction threshold, with the critical dynamics at this continuous active-to-absorbing state transition governed by the scaling exponents of directed percolation. Novel features in biologically motivated model variants include the stabilizing effect of a periodically varying carrying capacity that describes seasonally oscillating resource availability; enhanced mean species densities and local fluctuations caused by spatially varying reaction rates; and intriguing evolutionary dynamics emerging when variable interaction rates are affixed to individuals combined with trait inheritance to their offspring. The basic susceptible-infected-susceptible and susceptible-infected-recovered models for infectious disease spreading near their epidemic thresholds are respectively captured by the directed and dynamic isotropic percolation universality classes. Systems with three cyclically competing species akin to spatial rock-paper-scissors games may display striking spiral patterns, yet conservation laws can prevent such noise-induced structure formation. In diffusively coupled inhomogeneous settings, one may observe the stabilization of vulnerable ecologies prone to finite-size extinction or fixation due to immigration waves emanating from the interfaces.

2                                      *Uwe C. Täuber*

# Contents

## 1. Introduction: Stochastic spatial population dynamics

The study of living organisms, which per definition constitute complex dynamical systems comprising many interacting individual constituents, has become a vast and fertile research arena for physicists and mathematicians trained in nonequilibrium statistical mechanics and the theory of stochastic processes. Comparing with the more traditional applications of statistical physics to condensed matter systems, one may observe many common but also distinct features in biological applications. While the number of degrees of freedom for biological systems is certainly sufficiently large to render a statistical description meaningful and productive, it is also typically much smaller than Avogadro's number. Hence statistical fluctuations are usually non-neglible, become amplified owing to their being situated far away from thermal equilibrium, and are often driven by random external influences. Fluctuations are increasingly acknowledged to importantly contribute to biological functionality. Consequently the thermodynamic limit may not apply, merely studying average quantities is insufficient, finite-size effects are crucial, and transient phenomena dominate any living entity.

Biological structures as well as dynamical processes are also characteristically optimized due to evolutionary mechanisms. Thus both spatial and temporal correlations play prominent roles, manifesting themselves in spontaneous structure formation and the presence of history dependence and memory effects. In addition, as opposed to atoms or molecules, cells and individuals in a population are not identical, but carry specific traits that affect their dynamics and may also change over time. Nevertheless, as a theoretical physicist one hopes – almost in defiance of the astounding complexity of the biological realm, and despite the abundant difficulties in separating clearly distinct length and time scales that is so profoundly important in establishing well-defined effective description levels – that exhaustive studies of comparatively simple fundamental models that employ the full, powerful analytical and computational machinery developed in statistical physics may contribute a deeper understanding of both structural and dynamical features encountered in the living world.

This chapter addresses exemplary basic models in ecology and epidemiology that share a common mathematical language with chemical reactions, namely stochastic spatially extended systems that are dominated by a binary "predation" or "infection" process $A + B \to A + A$, where upon mutual encounter an individual or particle of "prey" or "susceptible" species B becomes replaced with a "predator" or "infectious" individual A, with a prescribed rate $\lambda$ that will serve as the pertinent nonlinear control parameter.[1–3] The corresponding mean-field rate equations are the celebrated coupled ordinary differential equations independently proposed about a century ago by Alfred Lotka and Vito Volterra who aimed to mathematically capture oscillatory dynamics in the quite distinct contexts of autocatalytic chemical reactions[4] and fish populations in the Adriatic sea.[5]

Indeed, much of the traditional original and textbook literature on chemical kinetics,[1,2] population ecology,[2,3,6,7] and closely related evolutionary game theory[8,9] invokes deterministic chemical rate equations or their spatial extension, reaction-diffusion models, which constitute mean-field type approximations that neglect stochastic fluctuations and spatio-temporal correlations through "mass action" factorizations of higher moments, see Sec. 2.[10–13] With the dramatic enhancement of computational power since the 1970s and the accompanying development of versatile analytical tools, stochastic models for (bio-)chemical reaction kinetics, population dynamics, and infectious spreading have become an increasingly active research field in nonequilibrium statistical physics, with the focus often on relevant fluctuation and correlation effects that are not captured by their representation in terms of mere rate equations[10–26] (this is merely a non-exhaustive list of reviews, recent monographs, and textbooks; please also see the extensive references cited therein).

As described in Sec. 3, Lotka–Volterra predator-prey systems display many non-trivial phenomena that are strongly influenced by intrinsic dynamical correlations and fluctuations: The species coexistence regime is dominated by persistent activity waves traversing the system where surviving predators necessarily pursue prey that expand into unoccupied space. These noise-generated and -stabilized correlated spatio-temporal structures in turn induce stochastic and resonantly amplified population oscillations whose quantitative features are markedly renormalized relative to mean-field theory. Through a nonlinear feedback mechanism, limited prey resources encoded in a finite local carrying capacity severely affect the predator population, even causing its extinction and thus prey fixation at a critical threshold $\lambda < \lambda_c$. The ensuing stationary critical dynamics as well as universal transient aging features that characterize the far-from-equilibrium continuous phase transition from the active coexistence to the inactive, absorbing predator extinction state are governed by the scaling exponents of the generic directed percolation universality class.

The biological context moreover suggests model variations that would not usually be considered in a chemical or physical setting; some exam-ples are discussed in Sec. 4, namely periodic changes in the environment and reaction rates that are specific to individual particles and whose distribution may evolve under Darwinian competition with mutations. Subsequently, Secs. 5 and 6 respectively provide succinct overviews of the stochastic spatial dynamics observed in the closely related fundamental susceptible-infected-susceptible and susceptible-infected-recovered models for the spreading of infectious diseases, specifically their near-threshold critical dynamics, as well as spontaneous spiral pattern formation or the absence thereof in the May–Leonard and "rock-paper-scissors" model variants designed to describe cyclic dominance of three competing species. The chapter concludes (Sec. 7) with the author's brief perspective and outlook.

## 2. Fluctuations and correlations in reacting particle systems

We begin with a concise outline of the mathematical description of stochastic reacting particle systems through master equations, highlighting the mean-field factorization approximation that is invoked to arrive at the associated coupled rate equations. The standard setup of Monte Carlo simulation algorithms to statistically sample observables is introduced as well.

### 2.1. *Stochastic kinetics and master equation*

"Chemical reactions" denote (stochastic) processes where through mutual interactions particles representing molecules of certain chemical species, nuclei or elementary particles, individuals in a population, etc. alter their identity. For example, consider the reversible binary reaction

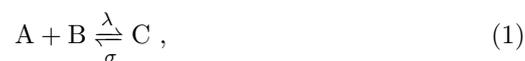

$$A + B \underset{\sigma}{\overset{\lambda}{\rightleftharpoons}} C \ , \tag{1}$$

where the notation "A + B" on the left-hand side implicitly expresses the condition that one particle each of species A and B must meet at the same location and time in order for the "forward" reaction with rate $\lambda$ to occur; in contrast, the reverse reaction with rate $\sigma$ may happen spontaneously as long as any C particles remain present. The configuration of the system at any time $t$ is fully characterized by specifying the (local) number of particles $n_\alpha = 0, 1, \dots$ of each species $\alpha = A, B, C, \dots$ at that instant. Since reactions just modify these integers $\{n_\alpha\}$ by the associated stoichiometric coefficients, they represent Markovian stochastic processes in species number space. In example (1), the reaction rates $\lambda$ and $\sigma$ then determine the associated configuration-dependent transition rates

$$w(\{n_A, n_B, n_C\} \to \{n_A - 1, n_B - 1, n_C + 1\}) = \lambda\, n_A n_B \;,$$
$$w(\{n_A, n_B, n_C\} \to \{n_A + 1, n_B + 1, n_C - 1\}) = \sigma\, n_C \;, \qquad (2)$$

as there are $n_\alpha$ possibilities to pick one of the reactant particles of type $\alpha$.

To generalize the above local processes to spatially extended systems, one merely needs to view each species $\alpha$ on distinct lattice or continuum positions $i$ as "quasi-species" that are fully characterized by integer occupation numbers $n_{\alpha\, i}$ which are in turn altered by integer values through reactions and / or transport. Unrestricted hopping of a single particle from site $i$ to $j$ with rate $D$ is then governed by the Markovian transition rate

$$w(\{n_{\alpha\, i}, n_{\alpha\, j}\} \to \{n_{\alpha\, i} - 1, n_{\alpha\, j} + 1\}) = D\, n_{\alpha\, i} \;. \qquad (3)$$

Yet if, for example, at most only a single particle may occupy any location $i$, so that all $n_{\alpha\, i} = 0$ or $1$, the right-hand side would need to be multiplied with the factor $(1 - n_{\alpha\, j})$, rendering the resulting exclusion process non-linear in nature. In biology, site restrictions that constrain local particle occupations may be caused by limited resources for the affected species, and the maximum sustained number is referred to as "carrying capacity".[2,3]

The "chemical" master equation constitutes a deterministic evolution equation for the configurational probability $P(\{n_\alpha\}; t)$ that accounts for the gain-loss balance due to reactive transitions,[10,27,28]

$$\frac{\partial P(\{n_\alpha\}; t)}{\partial t} = \sum_{\{n'_\alpha \neq n_\alpha\}} \Big[ P(\{n'_\alpha\}; t)\, w(\{n'_\alpha\} \to \{n_\alpha\}; t) $$
$$- P(\{n_\alpha\}; t)\, w(\{n_\alpha\} \to \{n'_\alpha\}; t) \Big] \;. \qquad (4)$$

In our example (1) governed by the rates (2),

$$\frac{\partial P(n_A, n_B, n_C; t)}{\partial t} = \lambda\, (n_A + 1)\, (n_B + 1)\, P(n_A + 1, n_B + 1, n_C - 1; t)$$
$$+ \sigma\, (n_C + 1)\, P(n_A - 1, n_B - 1, n_C + 1; t)$$
$$- (\lambda\, n_A n_B + \sigma n_C)\, P(n_A, n_B, n_C; t) \;, \qquad (5)$$

with the convention that $P(\{n_\alpha\}; t) = 0$ if any $n_\alpha < 0$. In terms of the varying particle numbers, Eqs. (5) represent an infinite set of coupled linear ordinary differential equations. Any observable quantity $O$ needs to be a function of the particle numbers $\{n_\alpha\}$, and hence its expectation value with respect to the ensemble of possible configurations is

$$\langle O(\{n_\alpha\}, t) \rangle = \sum_{\{n_\alpha\}} O(\{n_\alpha\})\, P(\{n_\alpha\}; t) \;, \qquad (6)$$

with its temporal evolution instilled by the master equation (4) for the probabilities $P(\{n_\alpha\}; t)$, which in principle allows us to compute characteristic time-dependent expectation values, moments, and correlation functions.

## 2.2. *Mean-field rate equations*

To illustrate this process, consider the average particle numbers,

$$\frac{\partial \langle n_\alpha(t) \rangle}{\partial t} = \sum_{\{n_\alpha\}} n_\alpha \frac{\partial P(\{n_\alpha\}; t)}{\partial t} \ . \tag{7}$$

As the infinite particle number sums range over all allowed integers, one may simply shift internal summation indices in the master equation gain terms to rewrite them as multiplying $P(\{n_\alpha\}; t)$; after resulting cancellations, the remainder can be expressed as other expectation values or moments.[10,11,13] For example, with Eq. (5) one arrives at

$$R(t) = \frac{\partial \langle n_{A/B}(t) \rangle}{\partial t} = -\frac{\partial \langle n_C(t) \rangle}{\partial t} = -\lambda \left\langle [n_A n_B](t) \right\rangle + \sigma \left\langle n_C(t) \right\rangle \ , \tag{8}$$

where the identities for the net reaction rate $R(t)$ on the left-hand side follow directly from the scheme (1); in general stoichiometric coefficients for the reactions would enter here. In the asymptotic long-time steady state, the average particle numbers become constant, and the exact relationship $\langle n_C(\infty) \rangle / \langle [n_A n_B](\infty) \rangle = \lambda / \sigma$ holds. Yet, in order to compute $\langle n_\alpha(t) \rangle$ for a reaction scheme involving binary or higher-order reactions, one requires the time evolution of nonlinear moments, which iteratively generates an infinite hierarchy of linear coupled differential equations for all moments or correlators. Closure towards a finite set of equations can only be achieved by assuming certain high- order correlations to be insignificant and applying factorizations of the associated moments in products of lower-order ones.[12]

The simplest, but also most drastic approximation is to neglect any two-point and higher- order correlations, which in our above example amounts to the mean-field decoupling $\left\langle [n_A n_B](t) \right\rangle \approx \langle n_A(t) \rangle \langle n_B(t) \rangle$ that should be applicable when the reactants are maintained well-mixed in the system, or for reactions in dilute gases or solutions with abundant particle numbers where any spatial and temporal fluctuations are very small compared to the mean reactant densities. Consequently, the reaction rate (8) reduces to $R(t) \approx -\lambda \langle n_A(t) \rangle \langle n_B(t) \rangle + \sigma \langle n_C(t) \rangle$, resulting in a closed set of three coupled (two of which are identical) but now nonlinear ordinary differential equations for the mean particle numbers. These rate equations may subsequently be analyzed by the standard methods of nonlinear dynamics.[1,2] For the stationary particle number averages, the mean-field approximation yields the ratio $\langle n_A(\infty) \rangle \langle n_B(\infty) \rangle / \langle n_C(\infty) \rangle = \sigma / \lambda$ which is just the chemical "law of mass action" applied to the reversible reactions (1) in the dilute reactant limit; inthermal equilibrium at temperature $T$, the logarithm of this backward-to-forward reaction rate ratio (the "pK value") can further-more be identified as the net reaction enthalpy relative to $k_{\mathrm{B}}T$.[28,29] In order to incorporate spatial variations, one may supplement the rate equations by, say, diffusive spreading terms $\sim D\,\nabla^2 n_\alpha(x,t)$ for the coarse-grained lo-cal particle densities, which generates "reaction-diffusion" equations that are however still subject to the mass action factorization assumptions.[2]

More refined mean-field theories apply the factorization approximation to higher-order correlations.[18,22] A systematic approach to faifhfully in-clude "demographic" fluctuations or "internal reaction noise" and corre-lations as well as to properly account for the fundamentally discrete nature of the stochastic particle interactions rests on a bosonic Fock state repre-sentation of the basic linear master equation kinetics and a subsequent con-struction of coherent-state path integrals in the continuum limit.[10,19,30–33]

The resulting Doi–Peliti field theory may subsequently be analyzed per-turbatively, with the associated "classical field equations" recovering the mean-field reaction-diffusion equations.[34,35] For at most binary reactions, the field theory action can be "reversely" mapped to stochastic Langevin partial differential equations that incorporate multiplicative noise.[10,34–38] Moreover, universal scale-invariant features can be extracted by means of dynamic renormalization group methods[10,19,25] (please also refer to the original citations therein).

## 2.3. *Individual-based Monte Carlo simulations*

An efficient numerical tool to investigate fluctuation and correlation effects in (bio-)chemical and physical reactions or ecological and epidemiologial dynamics is to implement stochastic "individual-" or "agent-based" Monte Carlo simulations.[13] These represent an underlying Markov chain master equation, since any configurational updates only depend on the immediate past, with the corresponding transition rates prescribed in the same man-ner as in the examples (2) and (3). Stochastic reaction, hopping, or other spreading processes can be implemented on regular lattices, often with pe-riodic boundary to eliminate boundary effects, or appropriately selected contact networks. A typical simulation algorithm proceeds as follows: The system is initialized by distributing particles of either species among available sites, typically randomly, but constrained by specified local occu-pation number restrictions, if applicable. Next, an individual is selected at random, and depending on the presence of other particles on its site $i$ or in its vicinity, possible reactions are executed with their specified probabilities and subject to any pertinent conditions. For example, pair processes require two particles to meet at the same or neighboring sites; and offspring production to an adjacent location may be restricted by a finite local carrying capacity. Note that in exclusion models with $n_{\alpha\,i} = 0$ or $1$ only, these reactions must necessarily be implemented "off-site", and birth processes hence inevitably generate spatial spreading even in the absence of explicit particle hopping or exchange processes. The particle selection and reaction steps are then repeated $N = \sum_{\alpha,i} n_{\alpha\,i}$ times such that on average each individual present in the system at this instant has been subject to one of the possible stochastic processes; this completes one Monte Carlo step (MCS) in the sequence of random updates.

Physical time is assumed to be proportional to this artificial simulation clock time. The algorithm may be run sufficiently long to reach a (quasi-)stationary configuration. To control statistical errors, adequate averages must subsequently be taken over many independent initializations and stochastic histories; these may be interpreted as direct sampling of observables subject to the time-dependent configurational probabilities evolving according to the corresponding master equation. Examples are the temporal evolution of mean particle numbers or appropriate correlations that provide quantitative mathematical insights; individual runs may be visualized as illustrative simulation movies. Properly sampled Fourier transforms of the time tracks may be utilized to detect characteristic frequencies and attenuation time scales of associated oscillations through Fourier spectrum peak locations and widths.

Of course, there are inevitable artifacts that manifest themselves on short time and length scales, originating in somewhat arbitrary choices in the algorithmic setups.[13] Yet the goal is to capture generic, perhaps emergent dynamical features on meso- or macroscopic coarse-grained scales that do not crucially depend on microscopic details. One should thus in principle execute distinct algorithms, and also check for the prominence of finite-size effects, e.g., when emerging dynamical structures reach across the entire system. It is also worth noting that in any case, macroscopic system parameters such as effective reaction rates are usually not directly and simply connected with the microscopic transition probabilities implemented in the computer code but may acquire nontrivial renormalizations due to intrinsic dynamical correlations.[39]

## 3. Predator-prey competition and coexistence

This section discusses the simplest stochastic spatially extended model for predator-prey competition and coexistence which reduces to the famed Lotka–Volterra equations in the mean-field approximation. Predators A and prey B by themselves are subject to linear death $A \xrightarrow{\mu} \emptyset$ and birth (asexual reproduction) processes $B \xrightarrow{\sigma} B + B$. Here, one may also interpret the parameters $-\mu$ and $\sigma$ as the net reproduction (difference of birth and death) rates for either species, which would respectively result in exponential decay $\langle n_A(t) \rangle = \langle n_A(0) \rangle e^{-\mu t}$ and growth $\langle n_B(t) \rangle = \langle n_B(0) \rangle e^{\sigma t}$. Competition and population control is achieved through the binary predation reaction $A + B \xrightarrow{\lambda} A + A$: Upon mutual encounter, a prey individual becomes replaced with a predator with rate $\lambda$.

### 3.1. *Lotka–Volterra model and population oscillations*

The master equation associated with the above stochastic predator death, prey birth, and predation processes reads

$$\frac{\partial P(n_A, n_B; t)}{\partial t} = \mu \, (n_A + 1) \, P(n_A + 1, n_B; t) + \sigma \, (n_B - 1) \, P(n_A, n_B - 1; t)$$
$$+ \lambda \, (n_A - 1) \, (n_B + 1) \, P(n_A - 1, n_B + 1; t)$$
$$- (\mu \, n_A + \sigma \, n_B + \lambda \, n_A n_B) \, P(n_A, n_B; t) \; . \tag{9}$$

Applying the mass action factorization approximation $\langle [n_A n_B](t) \rangle \approx \langle n_A(t) \rangle \langle n_B(t) \rangle$ then yields the original deterministic Lotka–Volterra model in the form of two coupled ordinary differential equations,

$$\frac{\partial \langle n_A(t) \rangle}{\partial t} = -\mu \, \langle n_A(t) \rangle + \lambda \, \langle n_A(t) \rangle \, \langle n_B(t) \rangle \; ,$$
$$\frac{\partial \langle n_B(t) \rangle}{\partial t} = \sigma \, \langle n_B(t) \rangle - \lambda \, \langle n_A(t) \rangle \, \langle n_B(t) \rangle \; . \tag{10}$$

These mean-field rate equations permit three distinct stationary solutions: (i) Total population extinction $\langle n_A(\infty) \rangle = 0 = \langle n_B(\infty) \rangle$; (ii) another absorbing state with predator extinction $\langle n_A(\infty) \rangle = 0$, but Malthusian prey population explosion $\langle n_B(t) \rangle \to \infty$; and (iii) predator-prey coexistence with finite $\langle n_A(\infty) \rangle = \sigma/\lambda$, $\langle n_B(\infty) \rangle = \mu/\lambda$. The predators of course benefit from high prey fertility $\sigma$, whereas the prey strive under conditions of high predator mortality $\mu$ and low predation efficacy $\lambda$; yet perhaps counterintuitively, increased $\lambda$ values cause a reduction in the predator population, too, due to a nonlinear feedback mechanism: Frequent predation events deplete the prey, leaving only scarce food recources for the predators.

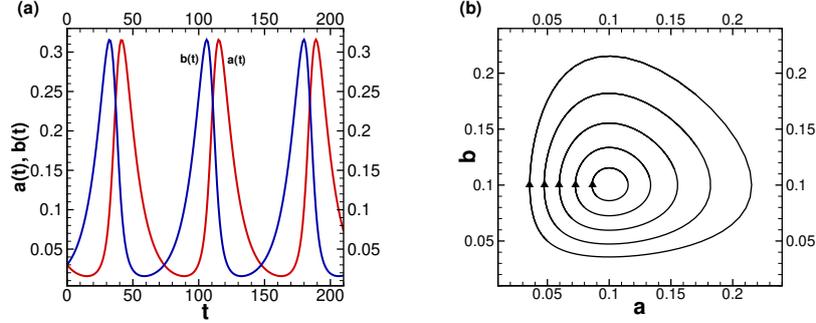

Fig. 1. (a) Predator (red) and prey (blue) population oscillations resulting from the deterministic Lotka–Volterra rate equations (10): $\sigma = 0.1$, $\mu = 0.1$, $\lambda = 1$. (b) Periodic orbits in population number phase space. [Figures reproduced with permission from Ref. [53], copyright (2007) by IOP Publ.]

The predator-prey coexistence state (iii) is however never reached under the mean-field dynamics (10) if the initial configuration is even slightly displaced from it: For small deviations $\delta_{A/B}(t) = \langle n_{A/B}(t) \rangle - \langle n_{A/B}(\infty) \rangle$, one may linearize the equations of motion to obtain $\partial \delta_A(t)/\partial t \approx \sigma \, \delta_B(t)$, $\partial \delta_B(t)/\partial t \approx -\mu \, \delta_A(t)$, which combine to the harmonic oscillator differential equation $\partial^2 \delta_{A/B}(t)/\partial t^2 \approx -\mu \, \sigma \, \delta_{A/B}(t)$ with the oscillation frequency $\omega_0 = \sqrt{\mu \, \sigma}$. Equivalently, one may construct the linear stability matrix

$$\mathbf{L} = \begin{pmatrix} 0 & \sigma \\ -\mu & 0 \end{pmatrix}$$

that dictates the time evolution $\partial \mathbf{v}(t)/\partial t \approx \mathbf{L} \, \mathbf{v}(t)$ of the fluctuation vector $\mathbf{v} = (\delta_A \; \delta_B)^T$. The eigenvalues $\pm i \omega_0$ of $\mathbf{L}$ are purely imaginary, indicating undamped oscillatory kinetics in phase space about the fixed point (iii). These neutral cycles are indeed confirmed via eliminating time from Eqs. (10): $d\langle n_A \rangle / d\langle n_B \rangle = [(\lambda \langle n_B \rangle - \mu) \langle n_A \rangle] / [(\sigma - \lambda \langle n_A \rangle) \langle n_B \rangle]$, which yields, after variable separation and integration, the conserved first integral (Lyapunov function) for the mean-field dynamics $l(t) = \lambda \langle n_A(t) \rangle + \lambda \langle n_B(t) \rangle - \sigma \ln \langle n_A(t) \rangle - \mu \ln \langle n_B(t) \rangle = l(0)$. Consequently, the trajectories in the population number phase space must be periodic orbits, shown in Fig. 1, that inevitably return to the initial configuration $\langle n_A(0) \rangle$, $\langle n_B(0) \rangle$.

However, the conservation of the quantity $l$ only holds within mean-field theory, and is not valid for the original stochastic reaction processes. Moreover, the fine-tuned vanishing real parts in the stability eigenvalues indicate a non-generic approximation artifact that should not remain robust as the model itself is altered. For example, one may implement a finite prey carrying capacity $K$ to prevent their population to diverge in case the predators go extinct. On the rate equation description level, this is captured by a logistic modification of the prey reproduction term,

$$\frac{\partial \langle n_B(t)\rangle}{\partial t} = \sigma \, \langle n_B(t)\rangle \left[1 - \frac{\langle n_B(t)\rangle}{K}\right] - \lambda \, \langle n_A(t)\rangle \langle n_B(t)\rangle \,. \qquad (11)$$

The ensuing modified stationary states are (ii') $\langle n_A(\infty)\rangle = 0$, $\langle n_B(\infty)\rangle = K$ and (iii') $\langle n_A(\infty)\rangle = (\sigma/\lambda)\,[1 - \mu/(\lambda K)]$, $\langle n_B(\infty)\rangle = \mu/\lambda$, which exists for $\lambda > \lambda_c = \mu/K$ and is then linearly stable. For $\lambda < \lambda_c$, the predator species is driven to extinction (ii'), while the prey population fixates at the carrying capacity $K$. At the coexistence fixed point (iii'), the eigenvalues of the stability matrix

$$\mathbf{L} = \begin{pmatrix} 0 & \sigma\,(1 - \mu/\lambda\,K) \\ -\mu & -\mu\,\sigma/\lambda\,K \end{pmatrix}$$

acquire negative real parts:

$$\epsilon_\pm = -\frac{\mu\,\sigma}{2\,\lambda\,K}\left[1 \pm \sqrt{1 - \frac{4\,\lambda\,K}{\sigma}\left(\frac{\lambda\,K}{\mu} - 1\right)}\,\right]. \qquad (12)$$

For $\sigma \geq \bar\sigma = 4\,\lambda\,K\,(\lambda\,K/\mu - 1)$, both eigenvalues are real, indicating direct exponential relaxation towards the stable node (iii'); on the other hand, if $\sigma < \bar\sigma$, the phase space trajectories spiral inwards to reach the stable focus (iii'), with the imaginary part of $\epsilon_\pm$ yielding the frequency of the resulting damped population oscillations.

The well-mixed rate equations can be amended to allow for spatial structures by replacing the mean population numbers with local density fields $n_{A/B}(x,t)$ (where $\langle n_{A/B}\rangle = \langle n_{A/B}(x,t)\rangle$) and adding diffusion terms,

$$\frac{\partial n_A(x,t)}{\partial t} = D_A \nabla^2 n_A(x,t) - \mu\,n_A(x,t) + \lambda\,n_A(x,t)\,n_B(x,t) \,, \qquad (13)$$

$$\frac{\partial n_B(x,t)}{\partial t} = D_B \nabla^2 n_B(x,t) + \sigma\,n_B(x,t) - \frac{\sigma}{K}n_B(x,t)^2 - \lambda\,n_A(x,t)n_B(x,t) \,.$$

In one dimension, these coupled partial differential equations support traveling wave solutions describing predator fronts that originate in a region set in the coexistence state (iii') and invade prey-occupied space.[2]

### 3.2. *Stochastic lattice model: Noise-induced activity fronts*

Monte Carlo simulations on sufficiently large two- and three-dimensional regular lattices (with periodic boundary conditions) qualitatively confirm several salient features predicted by the mean-field rate equations, but also display marked differences.[21,26,40–54] As shown in Fig. 2(a), prominent population oscillations are observed in the predator-prey coexistence state. Yet their properties are actually independent of the initial configurations and densities, at variance with the neutral centers resulting from Eqs. (10); indeed the quantity $l(t)$ is not constant in time, and stochastic fluctuations induce small damping in the resulting oscillatory modes. The population oscillation amplitudes are seen to diminish with increasing system size.

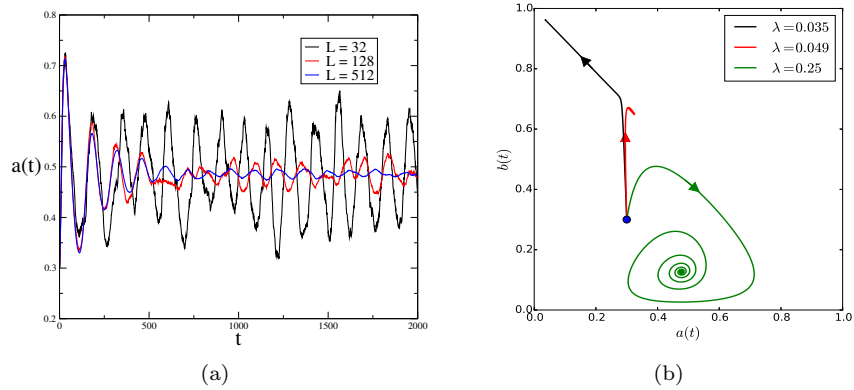

Fig. 2. (a) Predator population oscillations (single Monte Carlo simulation runs) in a stochastic Lotka–Volterra model with site exclusion ($n_{A/B\,i} = 0, 1$) on $L \times L$ square lattices (periodic boundary conditions) of linear system sizes $L = 32$, 128, and 512; initial predator and prey densities 0.3; $\mu = 0.025$, $\sigma = 1$, $\lambda = 0.25$; $D_{A/B} = 0$. The amplitude of the (damped) stochastic oscillations for the mean particle density decreases with $L$. (b) Typical phase space trajectories for single Monte Carlo runs on a $512 \times 512$ square lattice with $\mu = 0.025$, $\sigma = 1$, $D_{A/B} = 0$, and different predation rates, corresponding to the predator extinction / prey fixation state ($\lambda = 0.0035$), and two distinct predator-prey coexistence fixed points, either nodal ($\lambda = 0.049$) or focal ($\lambda = 0.25$). Near the latter, deep in the coexistence regime, one observes resonantly amplified oscillatory population fluctuations. [Figures reproduced with permission from (a) Ref. [21], copyright (2007) by Springer Science+Business Media, Inc.; (b) Ref. [26], copyright (2018) by IOP Publ.]

Upon implementing site occupation number restrictions, allowing at most a single particle of either species on each lattice point (i.e., $n_{A/B\,i} = 0, 1$), one observes three distinct regimes akin to the analysis of the Lotka–Volterra rate equations with finite prey carrying capacity Eq. (11), depicted in Fig. 2(b): Predator extinction and prey fixation for small predation rates $\lambda < \lambda_c$, albeit with a critical point location that is not in agreement with the mean-field prediction; a node-like two-species coexistence fixed point with exponential relaxation of the mean population numbers towards their steady-state values beyond the threshold $\lambda > \lambda_c$; and spiralling phase space trajectories deep in the coexistence regime for $\lambda \gg \lambda_c$ that reflect damped oscillatory kinetics, yet with manifest fluctuations near the focal center that are apparent as the broad "blob" in the (green) curve for $\lambda = 0.25$ in Fig. 2(b). Intriguingly, these fluctuations become more prominent with increasing distance from the active-to-absorbing state transition point.[21,26,52] For very large predation rates, even sizeable systems become vulnerable to total population extinction.[41,55] While any accessible absorbing state constitutes the ultimately only stable configuration in a finite system, the typical extinction times scale exponentially with the system size $L^d$, whence one typically observes species coexistence instead for sufficiently large $L$.[56,57]

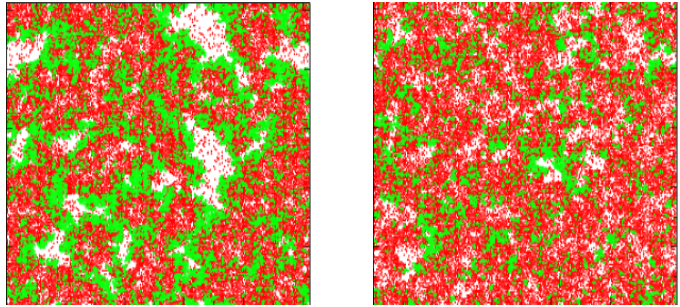

Fig. 3.   Monte Carlo simulation snapshots of a stochastic Lotka–Volterra model on a 256 × 256 square lattice subject to periodic boundary conditions and site exclusion. Predators (red) and prey (green) are initially randomly distributed; white: empty spaces. Pictured are the configurations at 500 (left) and 1000 MCS (right). [Figures reproduced with permission from Ref. [72], copyright (2016) by IOP Publ.]

Following the time evolution of individual simulation runs (Fig. 3) provides enlightening information on the origin of the erratic population oscillations and measured pertinent correlations in stochastic lattice Lotka–Volterra systems.[58] For large predation rates $\lambda$, deep in the two-species coexistence region, predators may initially devour almost all the prey present in the system. Subsequently, in the absence of food, their number decays rapidly with rate $\mu$ and the lattice may move close to total population extinction. Yet through the branching processes at rate $\sigma$, surviving prey B serve as randomly placed sources for invasion fronts that spread into empty space; these are then effectively "followed" by spared predators A, leaving

*Stochastic spatial Lotka–Volterra predator-prey models*        15

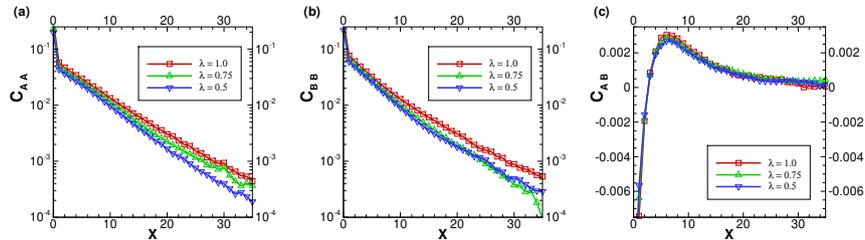

Fig. 4.   Equal-time (cross-)correlation functions (a) $C_{AA}(x)$, (b) $C_{BB}(x)$, and (c) $C_{AB}(x)$, inferred from Monte Carlo simulations of the site-unrestricted stochastic Lotka–Volterra model on a $1024 \times 1024$ square lattice in the (quasi-)stationary regime, with rates $\mu = \sigma = 0.1$, and different predation rates $\lambda = 0.5$ (blue), 0.75 (green), 1.0 (red). [Figures reproduced with permission from Ref. [53], copyright (2007) by IOP Publ.]

behind voids that may in turn be refilled by prey invididuals.[10,21,26,52–54] As depicted in the simulation snapshots in Fig. 3, the underlying stochasticity thereby generates intriguing persistent spatio-temporal structures that may be viewed as fluctuation-driven Turing patterns.[59,60] As these noise-stabilized evasion-pursuit waves traverse the system and interact, they generate stochastic population oscillations for both species. Interestingly, although correlations and cross-correlations remain short-ranged, with correlation lengths typically spanning just a few lattice spacings (Fig. 4), the associated local population fluctuations are quite large.[21,26,53,54,61] Bigger systems may however accommodate more out-of-phase activity fronts; consequently the global particle number oscillations decrease in amplitude.[21]

On one-dimensional lattices with multiple particles permitted on each site, the ballistically propagating evasion-pursuit fronts periodically wipe out large spatial regions;[53,54] if severe site restrictions are enforced, one instead observes segregation into prey and predator domains, which slowly coalesce and coarsen.[21] In spatial dimensions $d \geq 4$, the spontaneously generated spatio-temporal structures are not visible anymore, and the system's dynamics essentially attains mean-field character. This is in fact characteristic of two-species binary reactions, exemplified by simple pair annihilation $A + B \to \emptyset$, which is known to display species segregation and hence spontaneous cluster formation only in low dimensions $d \leq d_s = 4$.[10,11,25,62–65] In stark contrast to the Lotka–Volterra mean-field rate equations, the corresponding stochastic spatially extended system turns out remarkably robust with respect to model modifications.     For example, one may consider splitting the original binary reaction $A + B \to A + A$ perhaps more realis-tically into independent predation $A + B \rightarrow A$ and predator reproduction $A + B \rightarrow A + A + B$ processes, both requiring the presence of a prey individual adjacent to the predator. Whereas the mean-field scenario for the ensuing more complicated reaction scheme differs substantially from Eqs. (10), lattice Monte Carlo simulations for the altered stochastic reactions in fact yield the same qualitative behavior for both model variants.[52]

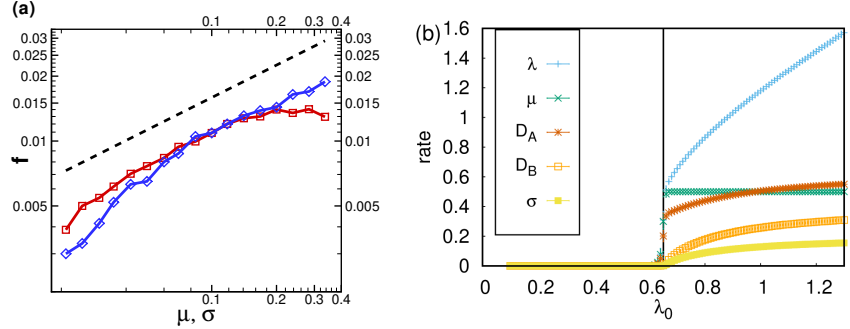

Fig. 5.  (a) Measured dependence of the characteristic peak frequency in the population number Fourier transforms $|\langle n_{A/B}(f)\rangle|$ on the rates $\sigma$ (red squares) and $\mu$ (blue diamonds), with $\lambda = 1$ and the respective other rate held fixed at the value 0.1, obtained from simulation data for a stochastic Lotka–Volterra system on $1024 \times 1024$ square lattices without site restrictions up to time $t = 20{,}000$ MC; for comparison, the dashed black line shows the (linear) mean-field oscillation frequency $f_0 = \sqrt{\mu\sigma}/2\pi$. (b) Macroscopic reaction rates $\mu$, $\sigma$, coupling $\lambda$, and diffusivities $D_{A/B}$ for the stochastic Lotka–Volterra predator-prey model on a square lattice with length $L = 150$, initial densities $a(0) = b(0) = 0.5$, as function of the microscopic predation rate $\lambda_0$, with fixed $\mu_0 = \sigma_0 = 0.5$, and $K = 1$; the vertical black line indicates the predator extinction / prey fixation threshold $\lambda_c$. [Figure (a) reproduced with permission from Ref. [53], copyright (2007) by IOP Publ.; (b) adapted from Ref. [39].]

### 3.3.  *Renormalized reaction rates and oscillation parameters*

Computing the Fourier transforms $\langle n_{A/B}(f)\rangle = \int \langle n_{A/B}(t)\rangle \, e^{2\pi i f t} \, dt$ of the population time tracks, the characteristic oscillation frequency may be extracted from the Fourier peak intensity, and the attenuation coefficient from its width. Results are displayed in Fig. 5(a), showing that the oscillation frequencies in the stochastic lattice Lotka–Volterra model are substantially reduced (by about $1/3$) relative to the linearized mean-field prediction. This indicates strong renormalizations due to fluctuations, but the data still approximately follow a square-root dependence on the rates $\mu$ and $\sigma$.[53] A recent study systematically explores the deviations of the measured coarse-grained reaction rates from the "microscopic" parameters for the computer simulation algorithm.[39] The dependences of the "macroscopic" predation coupling $\lambda$, predator death rate $\mu$ and prey birth rate $\sigma$, as well as both species' diffusivites $D_{A/B}$ on the input reactivity $\lambda_0$ are plotted in Fig. 5(b). As one should expect, the linear stochastic death processes occur independent of any spatial correlations, and hence $\mu$ remains unaltered until the predator extinction / prey fixation absorbing state transition at $\lambda_c$ is reached, beyond which of course all reactions cease. As the prey begin to fill the site-restricted square lattice with imposed local carrying capacity $K = 1$, prey proliferation and hopping become diminished and tend to zero as $\lambda_0 \to \lambda_c$, with constant ratio $\sigma/D_B$. In contrast, the predators' effective reaction rates jump discontinuously at $\lambda_c$.[39]

The Doi–Peliti formalism maps the stochastic master equation kinetics to a non-Hermitean bosonic many-particle Hamiltonian and ultimately employs coherent states, i.e., the complex eigenvalues of the annihilation operators, to construct a continuum field theory representation.[10,19,30–33] For the stochastic Lotka–Volterra predator-prey model with local site restrictions, the resulting action becomes[10,34]

$$S[\tilde{a}, \tilde{b}; a, b] = \int d^d x \int dt \left[ \tilde{a} \left( \frac{\partial}{\partial t} - D_A \nabla^2 + \mu \right) a + \tilde{b} \left( \frac{\partial}{\partial t} - D_B \nabla^2 - \sigma \right) b \right.$$
$$\left. - \sigma \, \tilde{b}^2 \, b + \frac{\sigma}{K} \left( 1 + \tilde{b} \right) \tilde{b} \, b^2 - \lambda \left( 1 + \tilde{a} \right) \left( \tilde{a} - \tilde{b} \right) a \, b \right] , \quad (14)$$

where the complex fields $a(x, t)$ and $b(x, t)$ originate from the coherent states, while the associated "response" fields are associated with their shifted adjoints, $\tilde{a} = \hat{a} - 1$, and similarly for $\tilde{b}$. The "classical" field equations associated with the action $\delta S/\delta a = 0 = \delta S/\delta b$ are solved by $\tilde{a} = 0 = \tilde{b}$, whereupon $\delta S/\delta \tilde{a} = 0 = \delta S/\delta \tilde{b}$ recovers the reaction-diffusion equations (13). Indeed, generally $\langle a(t) \rangle = \langle n_A(t) \rangle$ and $\langle b(t) \rangle = \langle n_B(t) \rangle$, but similar simple correspondences do not hold beyond the mean particle numbers.[10,19]

If one interprets the action (14) as the Janssen–De Dominicis functional in a path integral representation of stochastic partial differential equations,[36–38] it becomes equivalent to the coupled Langevin equations

$$\frac{\partial a(x, t)}{\partial t} = \left( D_A \nabla^2 - \mu \right) a(x, t) + \lambda \, a(x, t) \, b(x, t) + \zeta(x, t) , \quad (15)$$
$$\frac{\partial b(x, t)}{\partial t} = \left( D_B \nabla^2 + \sigma \right) b(x, t) - \frac{\sigma}{K} \, b(x, t)^2 - \lambda \, a(x, t) \, b(x, t) + \eta(x, t) ,$$

albeit for complex fields $a, b$.[10,34] These in effect add (complex) Gaussian stochastic forcings with zero mean $\langle \zeta \rangle = 0 = \langle \eta \rangle$ to Eqs. (13), subject to the noise (cross-)correlations

$$\langle \zeta(x,t)\, \zeta(x',t') \rangle = 2\lambda\, a(x,t)\, b(x,t)\, \delta(x - x')\, \delta(t - t')\ ,$$
$$\langle \zeta(x,t)\, \eta(x',t') \rangle = -\lambda\, a(x,t)\, b(x,t)\, \delta(x - x')\, \delta(t - t')\ ,$$
$$\langle \eta(x,t)\, \eta(x',t') \rangle = 2\sigma\, b(x,t)\left[ 1 - b(x,t)/K \right] \delta(x - x')\, \delta(t - t')\ ,\quad (16)$$

which entail multiplicative reaction noise terms that vanish as the particle densities approach zero, as is appropriate for the presence of a fully absorbing extinction state at $a = 0 = b$. Similar Langevin equations can be derived by means of a van Kampen system size expansion.[27,59]

In the predator-prey coexistence region, the mean population numbers $\langle a \rangle$ and $\langle b \rangle$ are finite. Consequently, the Langevin equations (15) describe spatially distributed nonlinear oscillators driven by additive white noise. The ensuing random kicks will occasionally be synchronous with the system's resonance frequency, causing large deviations from the coexistence fixed point densities, followed by attenuated population oscillations. This resonant amplification mechanism of internal stochastic fluctuations ultimately generates the persistent spatio-temporal structures and associated random oscillatory kinetics in Lotka–Volterra and related models.[10,21,34,66]

For a more detailed analysis, one may proceed by expanding the action in terms of fluctuating fields, defined relative to their mean stationary values. Diagonalization of the ensuing bilinear (Gaussian) action for equal predator and prey diffusivities $D_{\mathrm{A/B}} = D_0$ yields circularly polarized modes with dispersion relation $i\omega(q) = \pm i\omega_0 + \gamma_0 + D_0\, q^2$, where $\omega_0 = \sqrt{\mu\, \sigma\, (1 - \mu/\lambda\, K) - \gamma_0^2}$ is the "bare" oscillation frequency and $\gamma_0 = \mu\, \sigma/2\lambda\, K$ denotes the damping coefficient, see Eq. (12). In the absence of site occupation restrictions $K \to \infty$, $\gamma_0 \to 0$. Through a systematic perturbative calculation in terms of the effective expansion parameter $g = (\lambda/\omega_0)(\omega_0/D_0)^{d/2}$, one may evaluate fluctuation corrections to the oscillation frequency, diffusivity, and attenuation.[10,34] Although the parameter regime where perturbation theory is valid cannot easily be accessed in computer simulations, since the predation reactions would be too rare to generate decent statistics, the calculations confirm some of the pertinent numerical results: At least to first order in $g$, the characteristic oscillation frequency $\omega$ is shifted downward; this renormalization is particularly strong for $d \leq 2$ owing to the destructive interference of the two oppositely polarized eigenmodes which are almost massless due to the weak damping.

Also, the leading fluctuation corrections are symmetric in the rates $\sigma$, $\mu$ and are enhanced as $\sigma \ll \mu$ or $\sigma \gg \mu$; these features are in accord with the Monte Carlo data shown in Fig. 5(a). The diffusivity $D$ is shifted upwards by the fluctuations, resulting in faster propagation of the evasion-pursuit fronts, while the attenuation $\gamma$ is diminished compared to mean-field theory. Both these renormalizations help to induce instabilities towards spontaneous pattern formation that occur for $\gamma < 0$ at wavenumbers $q < \sqrt{|\gamma|/D}$.

### 3.4. *Predator extinction: Critical dynamics and aging*

Generically, the critical behavior at continuous nonequilibrium phase transitions from active to inactive, absorbing states is expected to be captured by the directed percolation universality class, at least in the absence of quenched disorder and any coupling to other conserved fields.[10,11,17–20,22,25,67,68] Since this Janssen–Grassberger conjecture also applies to reactive particle systems that incorporate multiple interacting species,[69,70] the observed dynamic scaling properties in stochastic spatial Lotka–Volterra models at the predator extinction threshold were immediately linked to critical directed percolation.[21,41,46,47,49,52] Heuristically, when near fixation the prey almost fill the entire system, the implicit condition for the predation reaction that one of the scarce predators must locate a prey individual is essentially always satisfied, whence the Lotka–Volterra processes reduce to the single-species (predator) reactions $A \to \emptyset$ and $A \rightleftharpoons A + A$, where the pair coagulation reaction originates from the site occupation restrictions or finite local carrying capacity.[21] These stochastic death, birth, and pair coagulation processes generate anisotropic directed percolation clusters in $(d+1)$-dimensional space-time, depicted in Fig. 14(a) in Sec. 5.1 below. When they all terminate after a finite time $t$, the system is in the absorbing state; when they extend to $t \to \infty$, it resides in the active phase. More formally, for $b \approx K$ the Doi–Peliti action (14) can be reduced to Reggeon field theory, which governs and in fact defines the directed percolation universality class.[10,11,21,25,26,34,67,71]

As Fig. 6 demonstrates, numerically extracting the dynamic scaling exponents from simulation data near the stationary regime at the stochastic Lotka–Volterra model's predator extinction / prey fixation phase transition does not produce satisfactory results even for fairly large lattices.[72] Indeed, double-logarithmic plots of the predator density decay, asymptotically to follow the power law $\langle n_A(t) \rangle \sim t^{-\alpha}$ at $\lambda = \lambda_c$, allow a precise determination of the critical point location (here at $\lambda_c = 0.0416$), independent of the lattice initialization. Yet a careful examination of the associated local slope or (time-dependent) effective decay exponent $\alpha_{\text{eff}}(t)$, displayed in the inset of Fig. 6(a), indicates that the universal asymptotic regime has barely been reached in these data, resulting in $\alpha \approx 0.54$; the accepted critical directed percolation value in two dimensions is $\alpha \approx 0.4505$.[10,22] Measuring the dynamic exponent $z\,\nu$ characterizing critical slowing down as obtained from analyzing the system's relaxation times $t_c \sim |\lambda - \lambda_c|^{-z\,\nu}$ following quenches near $\lambda_c$, depicted in Fig. 6(b), gives the satisfactory value $z\,\nu \approx 1.3$,[72] in accord with the literature exponents $z \approx 1.766$ and $\nu \approx 0.733$.[10,22]

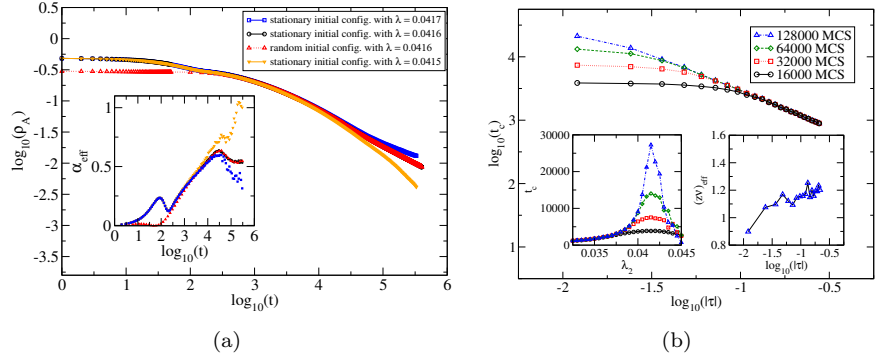

Fig. 6.  (a) Predator decay (double-logarithmic plots) for a stochastic Lotka–Volterra model on a $1024 \times 1024$ square lattice with $\mu = 0.025$ and $\sigma = 1$ at the prey fixation threshold $\lambda_c = 0.0416$, for quasi-stationary (top, black) and random (lower curve, red dotted) initial configurations (data averaged over 2000 independent simulation runs). For comparison, the density decay data are also plotted for $\lambda = 0.0417$ (blue, active coexistence phase) and $\lambda = 0.0415$ (orange, predator extinction regime). The inset shows the effective decay exponent $\alpha_{\text{eff}}(t)$, approaching $\alpha = 0.54 \pm 0.007$ at large $t$. (b) Characteristic relaxation time $t_c$ measured near the critical point. Left inset: $t_c(\lambda_2)$ after the system is quenched from a quasi-steady state at $\lambda_1 = 0.25$ to much smaller values $\lambda_2$ near $\lambda_c = 0.0416$, where the different graphs indicate $t_c$ when 128,000, 64,000, 32,000, and 16,000 MCS elapsed after the quench (data averaged over 500 runs). The main panel depicts these same data in double-logarithmic form; the graphs collapse for $|\tau| = |(\lambda_2/\lambda_c) - 1| > 0.1$, resulting in $z\,\nu = 1.208 \pm 0.167$. Right inset: associated effective exponent $(z\,\nu)_{\text{eff}}(\tau)$, approaching $z\,\nu \approx 1.3$ as $|\tau| \to 0$. [Figures reproduced with permission from Ref. [72], copyright (2016) by IOP Publ.]

Strong finite-size corrections may be avoided through seed dynamical Monte Carlo simulations, where initially a single active site, i.e., a predator A, is placed in the lattice, with the remainder occupied by prey B.[17,18] At the extinction critical point, the ensuing predator survival probability and density in this seed initial-slip regime should scale as $P_A(t) \sim t^\delta$ and $\langle n_A(t) \rangle \sim t^\theta$, where for directed percolation $\delta = \alpha = \beta/z\,\nu$ and $\theta = (2 - \eta)/z = (d/z) - 2\alpha$,[10,20] with $\theta \approx 0.2295$ in $d = 2$ dimensions.[10,22] Numerical analysis of the Monte Carlo simulation data displayed in Fig. 7 for a stochastic Lotka–Volterra model on a $512 \times 512$ square lattice with site restrictions yields the accurate critical exponents $\delta \approx 0.451$ and $\theta \approx 0.230$.[21]

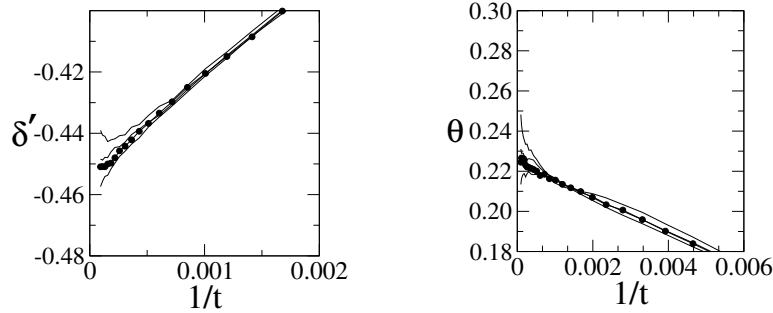

Fig. 7. Seed dynamical Monte Carlo simulations to estimate the predator extinction threshold $\lambda_c$ and critical exponents $\delta$ and $\nu$ for a stochastic site-restricted ($K = 1$) Lotka–Volterra model on a $512 \times 512$ square lattice with $\mu = 0.25$, $\sigma = 1$, $\lambda = 0.1690, 0.1689, 0.1688, 0.1687$ (top to bottom). The data are averaged over $3 \cdot 10^6$ independent runs with duration $10^5$ MCS. The effective critical initial slip exponents $\delta_{\text{eff}}(t)$ for the predators' survival probability $P_A(t)$ (left) and $\theta_{\text{eff}}(t)$ for their mean particle number $\langle n_A(t) \rangle$ (right) are plotted vs. $1/t$, giving $\lambda_c \approx 0.1688$. Extrapolation to $t \to \infty$ yields the estimates $\delta \approx 0.451$ and $\theta \approx 0.230$. [Figures reproduced with permission from Ref. [21], copyright (2007) by Springer Science+Business Media, Inc.]

Accessing critical dynamical scaling exponents through the universal short-time "physical aging" regime is another convenient means to avoid running costly simulations for very large system sizes $L$ that are required to meaningfully reach long times $t \sim L^z$.[10,73–76] To this end, one may for example suddenly switch the predation rate globally from $\lambda_1$ to a different value $\lambda_2$.[26,72] Outside the linear response regime, two-point (auto-)correlation functions $C(t, s) = \langle n(x, t)\, n(x, s) \rangle - \langle n(t) \rangle \langle n(s) \rangle$ will then depend on both involved times, often termed "waiting time" $s$ (after the initial system preparation) and "observation time" $t > s$. Only if the system's intrinsic relaxation processes are fast, typically exponential with characteristic time $t_c$, the preparation configuration prior to the quench is quickly forgotten, and time translation invariance recovered, as seen in the inset of Fig. 8(a), where both $\lambda_1, \lambda_2$ pertain to the coexistence phase.

In contrast, near a continuous phase transition, the relaxation time di-verges according to a power law. Consequently, for $t \gg s$ the two-time order parameter autocorrelation function obeys the scaling form[10,20,25,73,75,76]

$$C(t,s) = s^{-b} f_c(t/s) , \quad f_c(y) \sim y^{-\Lambda_c/z} , \tag{17}$$

which defines the aging scaling exponent $b$ and the autocorrelation exponent $\Lambda_c$.[76] For critical directed percolation the scaling relations $b = 2\alpha$ and $\Lambda_c/z = 1 + \alpha + (d/z)$ hold.[10,20,72] The numerical simulation data on a $1024 \times 1024$ square lattice with stochastic Lotka–Volterra kinetics demonstrates that time translation invariance is broken in quenches to the predator extinction / prey fixation threshold $\lambda_c$, c.f. Fig. 8(a), main panel. Instead, as demonstrated in Fig. 8(b), for sufficiently large waiting times $s \geq 1000$ MCS the aging scaling form (17) applies. The best data scaling collapse is obtained for $b \approx 0.88$, and the resulting slope of the master curve ultimately yields $\Lambda_c/z \approx 2.37$, in decent agreement with the established two-dimensional critical directed percolation exponents.[26,72] Intriguingly, the onset of shear turbulence in pipe flow displays spatio-temporal phenomena akin to predator-prey kinetics; hence its threshold properties are captured by the directed percolation universality class as well.[77] Indeed, the associated critical exponents have been most accurately measured ex-perimentally near the transition between two different turbulent states of electrohydrodynamic convection in thin nematic liquid crystals.[78,79]

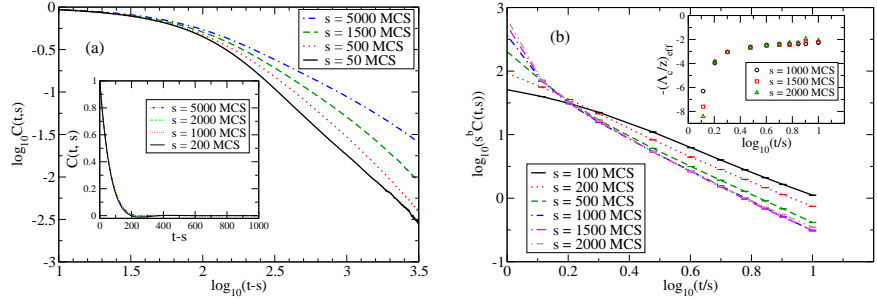

Fig. 8. (a) Double-logarithmic plot of the predator density autocorrelation function $C(t,s)$ vs. time difference $t-s$ for various waiting times $s=50, 500, 1500, 5000$ MCS (left to right) at the predator extinction critical point $\lambda_c = 0.0416$ for the same Lotka–Volterra system as in Fig. 6 (data averaged over 100 independent simulation runs for each value of $s$). The inset shows that $C(t,s)$ decays exponentially for $\lambda_2 = 0.125$, i.e., a quench within the coexistence phase; here time translation invariance holds (data averaged over 400 simulation runs for each $s = 5000, 2000, 1000, 200$ MCS, top to bottom). (b) Simple aging dynamical scaling analysis: $s^b C(t,s)$ is plotted against the time ratio $t/s$, with 1000 independent simulation runs for each $s$. The straight-slope section of the curves with large $s \geq 1000$ MCS yields $\Lambda_c/z = 2.37 \pm 0.19$; the aging scaling exponent is $b = 0.879 \pm 0.005$. The inset displays the local effective exponent $-(\Lambda_c/z)_{\mathrm{eff}}(t)$. [Figures reproduced with permission from Ref. [72], copyright (2016) by IOP Publ.]

## 4. Biologically motivated model variants and extensions

Naturally, a large variety of possible modifications and extensions can be envisioned for the simple Lotka–Volterra model for prey-predator competition and coexistence. Indeed, within the realm of coupled nonlinear mean-field rate equations, there exists a multitude of such generalizations studied in a vast literature encompassing different fields. Biological applications pertain to realizations in ecology, addressing predator-prey and host-parasite systems on all scales reaching from terrestrial mammals and fish populations to the microbiology of competing bacteria and oceanic plankton.

Stochastic spatially extended model implementations have been less frequently employed, but are becoming increasingly appreciated.[16] The complexity of multi-species ecological systems subject to many ill-defined external influences may be avoided in synthetic laboratory settings that are subject to controlled initial and environmental conditions. In that context, simple models in the spirit of the paradigmatic Lotka–Volterra system definitely constitute a powerful tool to both qualitatively and quantitatively describe experimental data; one example captures spatial range expansion of competing bacterial colonies, one of which may acquire resistance against toxins produced by the other species.[80]

On the theoretical front, the stability of multi-species systems and hence emergence of biodiversity remains an ecological puzzle, since within a mean-field framework, only the strongest predators or most evasive prey should survive direct competition. This scenario is however not confirmed in the corresponding stochastic model setups: For two distinct populations preyed upon by a single predator species, intrinsic demographic fluctuations are observed to prevent extinction of the weaker prey type.[81] Likewise, three-species coexistence in a system of two different predator groups competing for a single prey population may likewise be stabilized by stochasticity, especially when subject to evolutionary optimization.[82] In the following, just three biologically motivated Lotka–Volterra model variants are described in more detail, namely the system's response to periodic (seasonal) switches of the carrying capacity;[39] the effect of quenched random spatial disorder in the nonlinear predation rates mimicking environmental variability;[61] and the ensuing evolutionary dynamics upon introducing (in addition) individual demographic variability in the predation efficacy, whose distributions in both predator and prey species adjust themselves through the Darwinian mechanisms of competition and inheritance with random mutations.[26,54,83] More substantial variations of the Lotka–Volterra predation paradigm with prominent biological relevance are briefly discussed in Secs. 5 and 6 on stochastic models for epidemic outbreaks and cyclic competition.

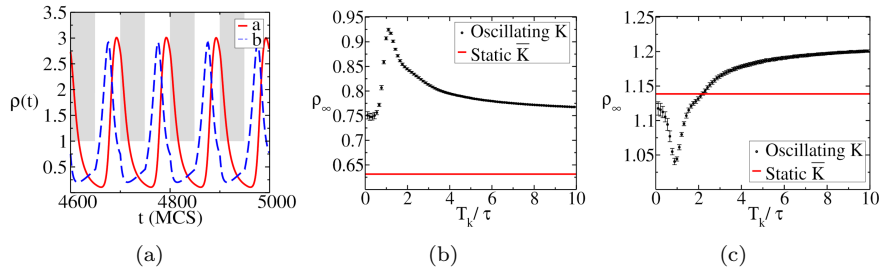

Fig. 9.   (a) Predator (full red) and prey (dashed blue) populations (averaged over 50 realizations) for a $256 \times 256$ square lattice with $\mu = \sigma = \lambda = 0.1$, $K_- = 1$, $K_+ = 10$, and $T_k = 100$; the shaded gray indicate the areas excluded by the switching carrying capacity $K(t)$. The critical predation rates associated with fixed carrying capacities are $\lambda_c(K = 1) = 0.26$ and $\lambda_c(K = 10) = 0.01$. Long-time (b) predator and (c) prey populations averaged over six periods $T_k$ plotted vs. $T_k/\tau$, where $\tau$ denotes the intrinsic oscillation period of the equivalent static system with $K = \bar{K}$; $L = 256$, $\mu = \sigma = \lambda = 0.1$, and $K_- = 2$, $K_+ = 6$ for the oscillating environment (black crosses), while $K = \bar{K} = 3$ for the static environment (red horizontal lines). [Figures reproduced with permission from Ref. [55], copyright (2023) by The American Physical Society.]

## 4.1. *Periodically varying carrying capacity*

Seasonal variations in resource availabilty, especially for the prey population, may be encoded in a periodically switching carrying capacity $K$; similarly, bacterial populations thriving on agar solutions in petri dishes may be supplied with nutrients at regular time intervals. In a stochastic Lotka–Volterra lattice model, one may directly implement a "rectangular time signal" for $K(t)$, switching from $K_-$ to $K_+ > K_-$ and back at full period $T_k$, in the simplest symmetric scenario remaining at $K_\pm$ for the duration $T_k/2$. If the entire carrying capacity range pertains to the species coexistence phase, only a minor amplification of the periodic population oscilations is observed. Conversely, predator extinction would of course ensue if $\lambda < \lambda_c(K_\pm)$. However, as shown in Fig. 9(a), the resulting time tracks for both predators and prey display large-amplitude periodicity if the carrying capacity oscillates between the predator extinction and species coexistence phases, i.e., in a static system with carrying capacity $K_-$, the A population would die out, but switching to $K_+$ above the critical threshold for a sufficiently long time allows the predators to recover.[55] For very fast switching, where $T_k$ is much smaller than the characteristic intrinsic population oscillation period $\tau$, the system becomes equivalent to an effective static model. Since in mean-field theory the stationary predator and prey populations $\sim 1/K$, their densities in the rapidly oscillating system are determined by the harmonic mean $\bar{K} = 2K_-K_+/(K_- + K_+)$. In the opposite extreme limit of very large $T_k \gg \tau$, both populations (in the coexistence phase) essentially merely switch between their stationary values, which is described by an equivalent static model with a rate-dependent carrying capacity $K^*$ that reduces to $\bar{K}$ if both $K_\pm \gg 1$.[55]

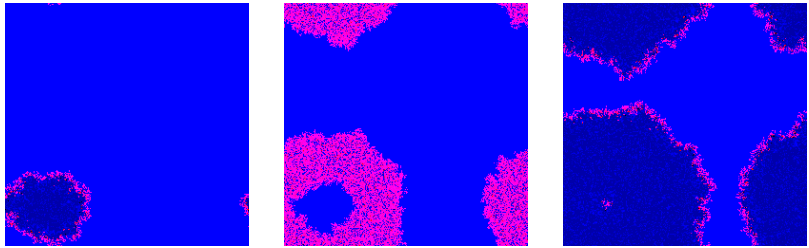

Fig. 10.   Simulation snapshots for a $256 \times 256$ square lattice with $\mu = \sigma = 0.5$, $\lambda = 0.1$, $K_- = 1$, $K_+ = 10$, and $T_k = 30$ MCS at $t = 39$ (left), 56 (middle), 72 MCS (right); red / blue pixels indicate predators / prey, where the brightness represents the local density, pink pixels pertain to sites occupied by both species; black: empty sites. The system is initialized with $K(t=0) = K_- = 1$. [Figures reproduced with permission from Ref. [55], copyright (2023) by The American Physical Society.]

Monte Carlo simulation results for stochastic lattice Lotka–Volterra models with periodically switching $K(t)$ confirm that the prey population is indeed generally quite well described by these effective static limits, within $\approx 10\%$ for the measured mean prey density, as seen in Fig. 9(c). Yet the mean predator population in Fig. 9(b) displays manifest quantitative deviations from the mean-field predictions, but of course approaches (slightly different) constant limits in both fast- and slow-switching regimes. Near $T_k \approx \tau$, intriguing resonances are visible. Related simulation snapshots are depicted in Fig. 10 for a switching period $T_k = 30$ MCS, close to resonance.[55] In this simulation run, by $t = 39$ MCS (left) only a single predator patch has survived, subsequently serving as the source for a spreading population front. At $t = 56$ MCS (middle) this wave expands in- and outward, until by $t = 72$ MCS (right) it returns to the source center; just then the system is switched back to the low carrying capacity $K_-$, whence the prey in the front interior cannot reproduce anymore. In its later temporal evolution, the system's kinetics continues to be dictated by the single population oscillation source generated by the sole predator patch surviving early on.

## 4.2. *Fitness enhancement through environmental variability*

Natural environments are of course not homogeneous, and subject to spatially varying resource availability, and specific to predator-prey interactions, habitat features that may locally enhance or suppress the chance of predation events. Such scenarios can be modeled through a spatially varying carrying capacity $K$, or site-dependent reaction rates, either smoothly varying or distinct in different regions or "patches". A perhaps extreme situation is to assign randomized rates to each discrete lattice site, picked for example from a Gaussian distribution, truncated for the reaction probabilities to the interval $[0, 1]$, and centered at a prescribed mean value with an assigned variance that indicates the "strength" for this quenched disorder.

As one would anticipate, varying rates $\mu, \sigma$ for the linear predator death and prey birth processes yield negligible changes to the mean populations, as random shifts just average out. That is not true if the predation rates are drawn from truncated Gaussian distributions with average $\bar{\lambda}$ and variance $\sigma_\lambda$, since (in the mean-field approximation) both stationary coexistence populations scale inversely with $\lambda$. Consequently, sites with low predation rates turn out favorable for, perhaps counterintuitively, both predator and prey species.[26,61] Indeed, as shown in Fig. 11(a), Monte Carlo simulations for two-dimensional Lotka–Volterra systems with quenched predation rate site disorder yield a $\sim 25\%$ "fitness" enhancement in the stationary densities of predators and prey as the width $\sigma_\lambda$ is increased towards an asymptotically flat distribution. Increasing disorder strength leads to the expected broadening of the associated Fourier peaks displayed in Fig. 11(b), which implies shorter relaxation times to reach the (quasi-)stationary configurations. Localization of the largest-amplitude local population oscillators near the favored sites with small $\lambda$ values additionally reduces the (cross-)correlation lengths by $\sim 1/3$ relative to the homogeneous system. Consequently more population centers can be accommodated by a system of fixed size, which explains the resulting enhancement in both predator and prey numbers.[26,61]

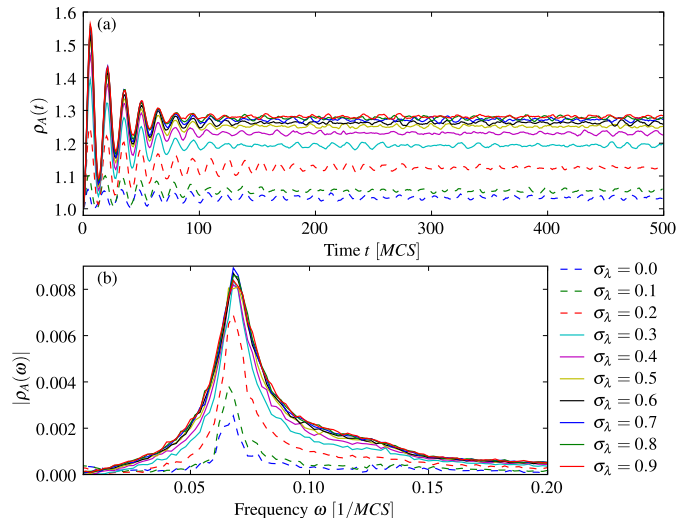

Fig. 11. (a) Predator population time evolution in a site-disordered Lotka–Volterra model on a $512 \times 512$ square lattice with $\mu = \sigma = 0.5$, mean predation rate $\bar{\lambda} = 0.5$, for different variances $\sigma_\lambda$ as indicated; data averaged over 50 independent simulation runs. (b) Fourier transform of the time traces. [Figure reproduced with permission from Ref. [61], copyright (2008) by The American Physical Society.]

### 4.3. *Demographic variability, heriditary trait optimization*

In contrast to physical and chemical reaction processes that involve fundamentally identical atoms, molecules, or similar microscopic constituents, individuals in a biological population are usually endowed with different propensities such as fecundity and efficacy in pursuing prey or evading predators. We may thus affix each reacting particle with, for example, a predation efficacy $\eta$. If a predator A and prey B encounter each other, we can for simplicity choose the resulting predation probability for this specific pair as the arithmetic mean $\lambda = (\eta_A + \eta_B) / 2$. These individual "hunting" or "evasion" efficacies can be turned into random variables drawn from, e.g., a Gaussian distribution with width $w_S$ truncated to the interval $[0, 1]$: Powerful predators are represented by $\eta_A \approx 1$, whereas evasion-proficient prey are characterized by $\eta_B \approx 0$. In this manner, genetic make-ups and behavioral features are subsumed into one "mesoscopic" phenotypical trait.[26,54,83]

This setup allows us to invoke rudimentary Darwinian evolutionary dynamics by allowing offspring to inherit their specific efficacy $\eta_O$ from their parent's individual $\eta_P$ value, with a random mutation "error". As Fig. 12 illustrates, in the (a) prey reproduction $P \to P + O$ at fixed rate $\sigma$ and (b) predation process $P + B \to P + O$ at variable rate $\lambda = (\eta_P + \eta_B)/2$, $\eta_P$ is set at the center of truncated Gaussian distribution with standard deviation (prior to truncation) $w_P$, from which $\eta_O$ is selected. The parameter $w_P$ describes the mutation strength; for $w_P = 0$ the offspring assumes precisely its parent's efficacy $\eta_P$. Since surviving prey will likely pass their small $\eta$ values to their progeny, while fertile predators will typically be endowed by large predation efficacies, the trait distributions for the two distinct populations will evolve in opposite directions.

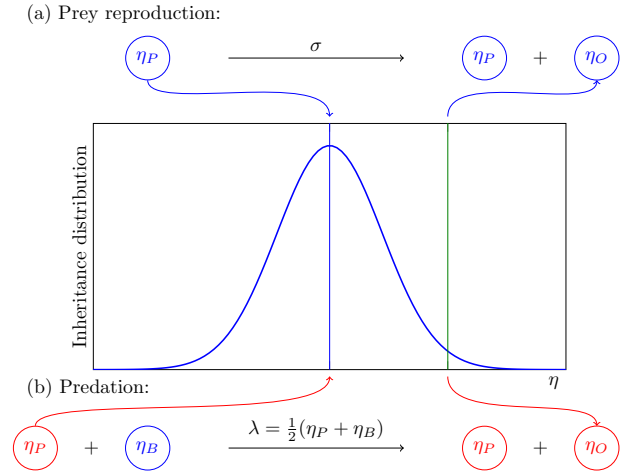

Fig. 12.  (a) During prey reproduction, a parent particle gives birth with rate $\sigma$. The parent's predation efficacy $\eta_P$ is set as the mean of a Gaussian distribution, truncated to the interval $[0; 1]$, from which the offspring's efficacy $\eta_O$ is drawn.  (b) During predation, a predator consumes a prey individual with rate $\lambda$, a function of the participating particles' predation efficacies.  An offspring predator is created and its efficacy $\eta_O$ is determined via the same mechanism as for prey reproduction.  [Figure reproduced with permission from Ref. [54], copyright (2013) by IOP Publ.]

Indeed, in Monte Carlo simulations of this model on a two-dimensional lattice both species' trait distributions settle, within a few hundred MCS (roughly equivalent to "generations"), to their stationary distributions shown in Fig. 13(a) and (b), respectively for a moderate mutation distribution width $w_P = 0.1$ and a flat distribution with $w_P \to \infty$. In the former scenario, the widths of the ensuing steady-state efficacy distributions for either species are comparable with $w_P$. In contrast with (linear) genetic drift models, this nonlinear evolutionary dynamics does not terminate in fixation of the prey at $\eta = 0$ and predators at $\eta = 1$; rather, demographic variabilities are sustained in both populations.[26,54,83] The dashed graphs indicate the numerical solutions of an effective mean-field model, wherein the traits $\eta$ are discretized and binned to define multiple "quasi-species" that are subject to the standard Lotka–Volterra rate equations. In the extreme case $w_P \to \infty$, the predators' selection bias disappears, and the stationary prey distribution $\sim 1/(1 + 2\eta)$ for the quasi-species model can be evaluated analytically, and decently captures the simulation data.

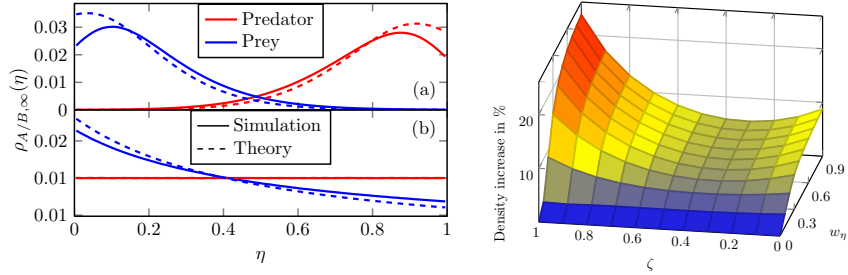

Fig. 13. Left panel: Predator (red) and prey (blue) populations in the (quasi-)steady state of a stochastic Lotka–Volterra model (no site occupation restrictions) on a $128 \times 128$ square lattice with $\mu = \sigma = 0.5$ and variable, hereditary predation efficacies $\eta$ affixed to all individuals, initially set to 0.5, as functions of $\eta$ (data averaged over $10^4$ Monte Carlo runs) for initial truncated Gaussian standard deviations $w_P = 0.1$ (a) and $w_P \to \infty$ (b). For finite $w_P$, the predator / prey distributions quickly evolve towards high / low predation efficacies. In the special flat distribution case, the predators experience no selection bias. The predictions from an effective stochastic multiple quasi-species mean-field model are depicted as dashed curves. Right panel: The stationary predator density features enhancement for all values of the spatial variability influence $\zeta$ as function of equal variabilities $w_W = w_D = w_\eta$ relative to uniform systems with $w_\eta = 0$; there appears a shallow minimum near $\zeta \approx 0.3$. [Figures reproduced with permission from Ref. [83], copyright (2013) by The American Physical Society.]

Finally, the combined effects of environmental and demographic variability in the predation processes can be studied.[26,54,83] To this end, random reaction reaction rates $\lambda_E$ attached to the lattice sites as well as predation efficacies $\eta_{A/B}$ relating to individual particles are implemented, each drawn from truncated Gaussian distributions with widths $w_E$ and $w_D$, respectively. In order to tune the relative influences of quenched spatial randomness and variabilities of the predator and prey populations, which are set to evolve in time according to the offspring hereditary selection with mutation chance described previously, we choose the linear interpolation $\lambda = \zeta \lambda_E + (1 - \zeta) \lambda_D$, where $\lambda_D = (\eta_A + \eta_B)/2$ as before, and $0 \le \zeta \le 1$.

The ensuing Monte Carlo simulation results for the stationary predator density are displayed in the right panel of Fig.13. For $\zeta = 1$, the $\sim 25\%$ fitness enhancement for exclusive site disorder described in Sec. 4.2 is recovered. In contrast, pure demographic variability $\zeta = 0$ increases the populations by less than 10%, indicating that the substantial evolutionary adjustments of both species' predation efficacy distributions almost cancel out in their competitive optimization leading to an almost neutral overall benefit. At $\zeta \approx 0.3$, an distinct minimum with even lower population enhancement is disernible; it can be understood in terms of straightforward error progression assuming statistical independence of environmental and demographical variability. The data displayed here demonstrate the dominance of environmental randomness. Yet in small systems the mean extinction time is increased more than fourfold in the presence of demographic variability which thus manifestly provides a major evolutionary advantage.[54,83]

## 5. Stochastic spatial models for infectious disease spreading

The basic Lotka–Volterra predation motif also appears in the fundamental mathematical models for epidemic outbreaks and spreading, dating back to Kermack, McKendrick, and Walker's seminal work from 1927.[2,84] Instead of predators and prey or parasites and hosts, one considers a population comprised of a now fixed total number $N$ of individuals that can be grouped into distinct states with respect to a contagious pathogen. To render these models more versatile and realistic, such "compartments" can be elaborated further by dividing them into additional distinct groups according to age or social structure, etc., and by implementing spatially extended stochastic models on appropriate contact network architectures.

### 5.1. SIS *model and directed percolation*

In the simplest scenario, healthy susceptible individuals S become infected with rate $r$ upon encountering sick or asymptomatic infectious agents I, following the binary (stochastic) reaction $S + I \xrightarrow{r} I + I$. The infected individuals may return to the susceptible group with recovery rate $a$, $I \xrightarrow{a} S$.[2,84] Note that these processes strictly conserve $N = n_S(t) + n_I(t)$; the latter transmutation may be regarded as number-conserving combination of the linear predator death and prey birth reactions of the Lotka–Volterra model,

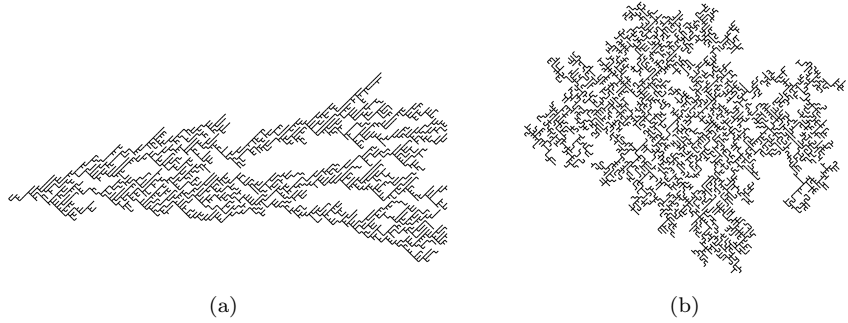

Fig. 14.  (a) Critical directed percolation cluster (time runs from left to right); (b) critical isotropic percolation cluster. [Figures reproduced with permission from Ref. [85], copyright (1994) by The American Physical Society.]

as is evident also in the corresponding mean-field rate equations

$$\frac{\partial \langle n_S(t) \rangle}{\partial t} = -r \langle n_S(t) \rangle \langle n_I(t) \rangle + a \langle n_I(t) \rangle \;,$$
$$\frac{\partial \langle n_I(t) \rangle}{\partial t} = r \langle n_S(t) \rangle \langle n_I(t) \rangle - a \langle n_I(t) \rangle \;. \tag{18}$$

Near the epidemic threshold, there are few infected individuals $n_I \ll n_S$; the S species is abundant and may thus be eliminated (integrated out). Thus, upon implementing any population-limiting mechanism such as lattice site restrictions, this "susceptible-infected-susceptible" (SIS) model near criticality reduces to the elementary death, birth, and coagulation reactions that generate directed percolation clusters; an example is shown in Fig. 14(a).

A continuum description for "simple epidemic processes" that encompass the SIS model can be formulated in terms of a Langevin stochastic partial differential equation for the local density of infectious individuals $n(x,t)$ that spread diffusively and are subject to reactions which require the presence of "active" agents,[10,17–20,22,25,67]

$$\frac{\partial n(x,t)}{\partial t} = \left(D\nabla^2 - \mathcal{R}[n(x,t)]\right) n(x,t) + \eta(x,t) \;. \tag{19}$$

In the threshold vicinity, i.e., near the continuous active-to-absorbing phase transition, the deterministic reaction kernel may be expanded according to $\mathcal{R}[n] \approx a + u\,n + \ldots$, which leads directly to the Fisher–Kolmogorov–Petrovsky–Piskunov equation.[2] The noise correlations too must obey the absorbing-state constraint and vanish as $n \to 0$. To leading order in the active density field, one arrives at the "square-root" multiplicative noise

$$\langle \eta(x,t)\,\eta(x',t') \rangle \approx v\,n(x,t)\,\delta(x-x')\,\delta(t-t') \;. \tag{20}$$

The Janssen–DeDominicis functional[10,36–38] associated with this Langevin kinetics (19), (20) is precisely the Reggeon field theory action.[10,19,20,71]

## 5.2. SIR *model and dynamic isotropic percolation*

After exposure to most infectious agents, individuals do not return to a susceptible state, but at least for some extended time period acquire immunity against renewed infection. Kermack, McKendrick, and Walker hence introduced a "recovered" compartment that in general incorporates both immune and deceased formerly infected populations; in this "susceptible-infected-recovered" (SIR) model the R species does not further partake in the system's dynamics. In a stochastic formulation, the associated reaction processes are $S + I \xrightarrow{r} I + I$ and $I \xrightarrow{a} R$. The total population number $N = n_S(t) + n_I(t) + n_R(t)$ remains fixed, but the infected fraction $n_I(t)$ ultimately becomes depleted as it enters the R compartment. From the associated rate equations

$$\frac{\partial \langle n_S(t) \rangle}{\partial t} = -r \langle n_S(t) \rangle \langle n_I(t) \rangle \ ,$$

$$\frac{\partial \langle n_I(t) \rangle}{\partial t} = r \langle n_S(t) \rangle \langle n_I(t) \rangle - a \langle n_I(t) \rangle \ ,$$

$$\frac{\partial \langle n_R(t) \rangle}{\partial t} = a \langle n_I(t) \rangle \ , \tag{21}$$

one obtains $d\langle n_S \rangle / d\langle n_R \rangle = -\langle n_S \rangle \, r/a$, which with $\langle n_R(0) \rangle = 0$ integrates to $\langle n_S(\infty) \rangle = \langle n_S(0) \rangle \, e^{-\langle n_R(\infty) \rangle \, r/a} \geq \langle n_S(0) \rangle \, e^{-N \, r/a} > 0$ and hence $\langle n_R(\infty) \rangle = a \int_0^\infty \langle n_I(t) \rangle \, dt = N - \langle n_S(\infty) \rangle < N$, since $\langle n_I(\infty) \rangle = 0$. Under the well-mixed mean-field assumption, $\langle n_I(t) \rangle$ will initially increase and thus induce an epidemic outbreak if $\langle n_S(0) \rangle > a/r$, or the (mean) "basic reproduction number" $\mathcal{R}_0 = \langle n_S(0) \rangle \, r/a > 1$; otherwise $\langle n_I(t) \rangle \to 0$ monotonically and the infection fizzles out.[2]

In spatially extended stochastic SIR models one observes distinct spreading infection fronts, similar to the activity waves in the Lotka–Volterra predator-prey system. Figure 15 compares the numerical solutions of the SIR rate equations for $\langle n_I(t) \rangle$ and $\langle n_R(t) \rangle$ with the corresponding mean populations extracted from individual-based Monte Carlo simulations for a stochastic model realization on a regular square lattice.[86] To this end, the effective "renormalized" rates (specifically the infectivity $r$) in the deterministic mean-field equations were adjusted to closely fit the peak intensity of the infection outbreak.    Clear deviations between the rate equation predictions and the simulation data that properly incorporate fluctuations and correlations are discernible both at the epidemic's onset and at its late stages:  Mean-field theory consistently overestimates $\langle n_I(t) \rangle$ (left panel), providing a too pessimistic outlook for controlling the disease outbreak such as enforcing regular testing protocols, quarantine of infected individuals, and isolation of their contacts.[87]  The deterministic mean-field equations yield an initial exponential growth for the infected population at rate $a\,(\mathcal{R}_0 - 1)$; the simulation data are rather described by power laws for both $\langle n_I(t) \rangle$ and $\langle n_R(t) \rangle$, as shown in the insets in Fig. 15.  Also, if the individuals are mobile (e.g., via nearest-neighbor hopping), the entire susceptible population becomes infected, so ultimately $\langle n_S(\infty) \rangle = \langle n_I(\infty) \rangle = 0$ and $\langle n_R(\infty) \rangle = N$.  The mean-field approximation then drastically underestimates $\langle n_R(\infty) \rangle$, by $\sim 15\%$ for the parameters chosen here (right panel).

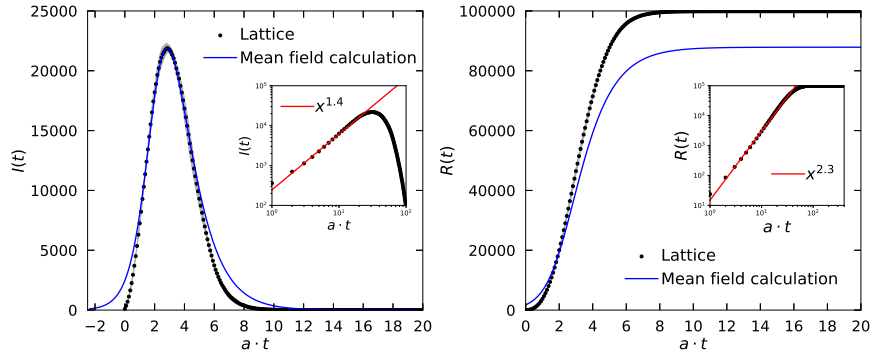

Fig. 15.  Individual-based Monte Carlo simulations for a stochastic SIR model on a $448 \times 448$ square lattice with $N = 100,000$ individuals, initially $n_I(0) = 100$, $n_R(0) = 0$, infectivity $r = 1$, mean recovery time $1/a = 6.667$ days, giving $\mathcal{R}_0 = 2.4$ adequate for the original Covid-19 viral strain.  The graphs show first of the infection curves from lattice simulations (averaged over 100 independent runs) to numerical integrations of the mean-field SIR rate equations.  Left: Infectious population $\langle n_I(t) \rangle$; right: recovered individuals $\langle n_R(t) \rangle$.  [Figure reproduced with permission from Ref. [86], copyright (2021) by Nature Research.]

The SIR model represents a wider class of "general epidemic processes" that entails the effective removal of (permanently) immune or deceased individuals, but also supposes that all previously infected agents I remain contagious to adjacent susceptibles S until they transition to the R class.  This persistent infectivity introduces temporal memory; in the corresponding continuum representation, integrating out the infectious agent field again yields a Langevin equation of the form (19) with multiplicative noise (20), but with a reaction kernel $\mathcal{R}[n] \approx a + u \int^t n(x, t')dt' + \ldots$ that contains the accumulated "debris". This defines the dynamic isotropic percolation universality class.[10,17,18,20,22,25,88–90] Indeed, for $t \to \infty$ the associated Janssen–De Dominicis functional reduces to the stationary effective action that governs critical isotropic percolation clusters, see Fig. 14(b). Numerical simulations confirm these assertions even in the presence of lattice site dilution, representing randomly distributed immunized individuals.[91]

## 6. Cyclic dominance of three competing species

In ecological "food networks", there exist also cyclic competition motifs in addition to hierarchical food chain structures.[6,7,26,92] Such cyclically implemented dominance akin to the popular "rock-paper-scissors" children's game plays a prominent role in evolutionary game theory as well,[8,9] and may be experimentally realized in microbial colonies.[23,24,26] We will next briefly discuss some pertinent features of the two probably most frequently studied model variants for three species cyclically preying on each other.

### 6.1. *Rock-paper-scissors or cyclic Lotka–Volterra model*

The simplest rock-paper-scissors model just cyclically combines the Lotka–Volterra predation reactions $A + B \xrightarrow{\lambda_A} A + A$, etc.[8,9] Importantly, the total particle number $N = n_A(t) + n_B(t) + n_C(t)$ is strictly conserved. If we allow the three particle species to propagate via hopping to (or position swaps with) adjacent sites, the corresponding continuous density field obeys the diffusion equation. Similar to the two-species predator-prey system with infinite carrying capacity, the associated mean-field rate equations

$$\frac{\partial \langle n_A(t) \rangle}{\partial t} = \langle n_A(t) \rangle \left[ \lambda_A \langle n_B(t) \rangle - \lambda_C \langle n_C(t) \rangle \right], \quad \text{(and cyclic)} \qquad (22)$$

feature the coexistence fixed point $\langle n_A(\infty) \rangle = N \lambda_B / (\lambda_A + \lambda_B + \lambda_C)$, etc., with linear stability matrix

$$\mathbf{L} = \frac{N}{\lambda_A + \lambda_B + \lambda_C} \begin{pmatrix} 0 & \lambda_A \lambda_B & -\lambda_B \lambda_C \\ -\lambda_A \lambda_C & 0 & \lambda_B \lambda_C \\ \lambda_A \lambda_C & -\lambda_A \lambda_B & 0 \end{pmatrix},$$

with eigenvalues 0 (for the conserved total particle number $N$) and $\pm i \omega_0$ describing neutral cycles and hence undamped population oscillations with frequency $\omega_0 = N \sqrt{\lambda_A \lambda_B \lambda_C} / (\lambda_A + \lambda_B + \lambda_C)$.

In a spatially extended stochastic cyclic Lotka–Volterra system, one might thus expect the emergence of activity waves akin to the two-species predator-prey model. However, these are not actually observed; instead, such rock-paper-scissors lattice models are characterized by mildly fluctuating particle clusters of the same species,[93,94] see Fig. 16(a). Evidently the diffusive mode associated with the total particle number conservation is quite effective in eliminating any larger-scale spatio-temporal structures. Moreover, the associated species number anti-correlations prevent synchronous local coherent oscillatory modes.[35] As stochastic noise generates a marked attenuation of the population oscillations, whose frequencies are slightly shifted downwards relative to the mean-prediction $\omega_0$, each species reaches its stationary density comparatively fast.[94] A perturbative analysis based on the corresponding Doi–Peliti action qualitiatively confirms these numerical findings. The fluctuation-induced damping coefficient $\gamma$ is positive, at least to first order in $\lambda/D$, which indicates that the spatially uniform species distribution remains stable.[35]

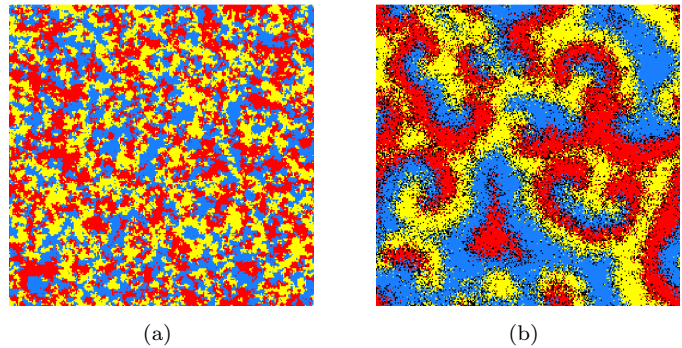

Fig. 16.   Monte Carlo simulation snapshots of stochastic cyclic competition models on $256 \times 256$ square lattices with periodic boundary conditions, site occupation number exclusion, and random initial particle placement: (a) Cyclic Lotka–Volterra (rock-paper-scissors) model with conserved total particle number $N$; (b) May–Leonard model with separate predation and reproduction processes. Sites occupied by A, B, and C particles are respectively colored in red, yellow, and blue; black: empty spaces. [Figures reproduced with permission from (a) Ref. [94], copyright (2010) by The American Physical Society; (b) Ref. [99], copyright (2011) by EDP Sciences.]

Intriguingly though, in the extreme asymmetric predation rate limit $\lambda_A \gg \lambda_B, \lambda_C$, the rock-paper-scissors system effectively recovers the two-species Lotka–Volterra model with predators A, prey B, and abundant C population that essentially fills the system, $\langle n_C(\infty) \rangle \approx N$.[35,95] On the mean-field level, neglecting fluctuations in the C density yields the effective predator death rate $\mu = \langle n_C(\infty) \rangle \lambda_C$ and prey birth rate $\sigma = \langle n_C(\infty) \rangle \lambda_B$.

A field-theoretical analysis confirms the mapping of the strongly asymmetric rock-paper-scissors Doi–Peliti action to its Lotka–Volterra counterpart (14), or equivalently to the effective Langevin equations (15) with noise correlations (16).[35]  Indeed, lattice simulations of the rock-paper-scissors model with massive asymmetry in the predation rates display the characteristic Lotka–Volterra evasion-pursuit fronts and persistent population oscillations.[95]

## 6.2. *Spiral structure formation in the May–Leonard model*

The cyclic three-species May–Leonard model probably more realistically separates the predation $\sim \lambda$ and reproduction reactions $\sim \sigma$ to independent stochastic processes,[92] which lifts the constraint of total population conservation and induces a much richer phenomenology than in the cyclic Lotka–Volterra model.[23,24,26]  It is defined by the competing reactions $A + B \xrightarrow{\lambda} A$ (and cyclic) and $A \underset{\kappa}{\overset{\sigma}{\rightleftharpoons}} A + A$, where the reverse reaction $\sim \kappa$ is set to limit the overall population, and where uniform rates across all three species have been assumed here.  The corresponding mean-field rate equations read

$$\frac{\partial n_A(t)}{\partial t} = \langle n_A(t) \rangle \left[ \sigma - \kappa \langle n_A(t) \rangle - \lambda \langle n_C(t) \rangle \right], \quad \text{(and cyclic)}, \qquad (23)$$

giving the symmetric three-species coexistence fixed point $\langle n_{A/B/C}(t) \rangle = \sigma / (\kappa + \lambda)$ with associated stability matrix

$$\mathbf{L} = -\frac{\sigma}{\kappa + \lambda} \begin{pmatrix} \kappa & 0 & \lambda \\ \lambda & \kappa & 0 \\ 0 & \lambda & \kappa \end{pmatrix}.$$

It has a negative real eigenvalue $-\sigma$ and the complex conjugate pair $-\sigma \left( 2\kappa - \lambda \pm i\sqrt{3}\,\lambda \right) / 2 \left( \kappa + \lambda \right)$, which describe oscillations at frequency $\omega_0 = \sqrt{3}\,\sigma\,\lambda / 2 \left( \kappa + \lambda \right)$ whose amplitude decays for $2\kappa > \lambda$, but blows up for $2\kappa < \lambda$, indicating a Hopf bifurcation at $\kappa_c = \lambda/2$. For $\kappa > \kappa_c$, the mean-field coexistence fixed point is a stable focus, and approached in spiralling trajectories.  Conversely in the unstable regime for $\kappa < \kappa_c$, the population trajectories in phase space assume the form of large-amplitude heteroclinic cycles, which in a stochastic system may reach the absorbing boundaries where either two species become extinct.  The likelihood of fixation is exacerbated in strongly asymmetric stochastic May–Leonard systems.[96]

In a sufficiently large spatially extended May–Leonard model realization, the spontaneous emergence of spiral structures as depicted in Fig. 16(b) renders the system remarkably robust against single-species fixation,[97,98] even when model variations such as quenched disorder in the rates are introduced.[99,100] Yet as with increased particle mobility the characteristic spiral arm diameter approaches the linear exxtension $L$, the May–Leonard kinetics is driven unstable in finite systems and becomes prone to two-species extinction.[97] Near the Hopf bifurcation at $\kappa_c$, critical slowing down of the circularly polarized damped oscillating modes relative to the much faster (at rate $\sigma$) relaxing total particle density field affords a natural time scale separation in the system,[98] allowing one to integrate out the fast variable and map the rate equation model to the complex Ginzburg–Landau equation, one of the prime paradigms for spontaneous structure formation out of equilibrium.[101] However, for the fully stochastic May–Leonard model, the range of validity for this mapping becomes quite limited; through the non-vanishing noise cross-correlations, the fast mode is continously excited, and one must resort to three stochastic fields to adequately capture the system's kinetics.[102] This can be achieved perturbatively via the associated Doi–Peliti path integral representation; yet the fluctuation corrections to lowest order do not alter the conclusions from the mean-field picture in a qualitative manner, but merely shift the Hopf bifurcation point and generate small oscillation parameter renormalizations.[35]

Generalization of the cyclic dominance motif to multiple competing species that may also form "alliances" depending on the structure of their interaction matrices has become a rich and intriguing research arena that can however not be adequately covered here.[26,103–114] It is worth mentioning though that an analysis of the associated adjacency matrix allows a remarkably complete classification of such many-species games with substructure within the mean-field theory realm, with decisive ramifications for the expected spontaneously forming patterns and ensuing coarsening kinetics in their spatially extended stochastic counterparts.[26,113]

### 6.3. *Inhomogeneous systems: diffusively coupled patches*

To conclude this discussion of stochastic spatial systems subject to binary predation processes, consider now spatially inhomogeneous settings where distinct "patches" controlled by different reaction rates are combined, and become effectively coupled through diffusive particle transport across their interfaces: As individuals cross the subsystem boundaries, they follow the stochastic processes and rules of their new environment. An example is illustrated in Fig. 17 for a series of successively finer-grained checkerboard squares of two-dimensional Lotka–Volterra predator prey patches alter-natingly situated in the predator extinction and two-species coexistence regimes.[115] In the quasi-steady state, each subsystem has clearly evolved to its stationary prey fixation (uniformly green) or mixed-population configurations. Predators from the coexisting patches enter the fixation regimes, but the range of these immigrations remains constrained to thin boundary layers of the extent of a few lattice sites, set by the correlation length in the fixated region. As the number of subdivisions increases and the individual patch sizes reach the underlying lattice constant, the measured mean population densities approach the corresponding results for a quenched random predation rate assignment drawn from a bimodal distribution.[115]

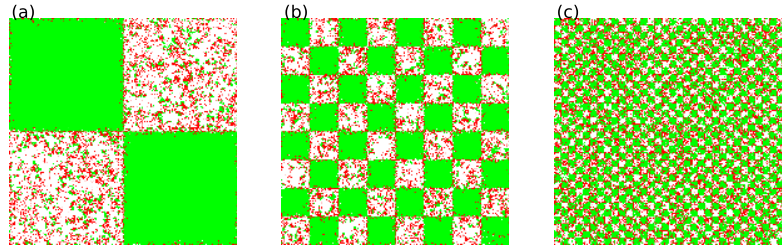

Fig. 17.  Simulation snapshots after 1000 MCS of the distribution of predators (red) and prey (green) in a stochastic Lotka–Volterra model on a spatially inhomogeneous $512 \times 512$ lattice (with periodic boundary conditions) with rates $\mu = 0.125$, $\sigma = 1$, and the predation rate set alternantly in a checkerboard fashion between square patches with values $\lambda = 0.1$ (prey fixation) and $\lambda = 0.8$ (species coexistence). The system is split into successively smaller subdomains with linear sizes 256 (a), 64 (b), 16 (c). [Figures reproduced with permission from Ref. [115], copyright (2017) by Elsevier B.V.]

Figure 18(a) schematically shows another realization of a diffusively coupled inhomogeneous setup, here for three-species cyclic rock-paper-scissors patches, with a thinner slab evolving under the cyclic Lotka–Volterra rules interacting with a much wider region governed by the May–Leonard reactions.[96,116] The small fluctuating clusters characteristic of the cyclic Lotka–Volterra model with conserved total particle number are visible on the left-hand side of Fig. 18(b). As the periodic population oscillations coherently impinge on the two interfaces, they generate planar invasion fronts moving into the larger May–Leonard region, which however decay on the scale of the characteristic correlation length $\xi$ in that regime. In the interior of the May–Leonard patch, therefore its typical spiralling spatio-temporal patterns form, largely decoupled and independent from any influences from the rock-paper-scissors domain.   In the vicinity of the interfaces, the population densities in the May–Leonard domain are reduced relative to the bulk, closer resembling the well-mixed mean-field values which do not account for the fitness enhancement due to the emergence of spiral structures.[116]

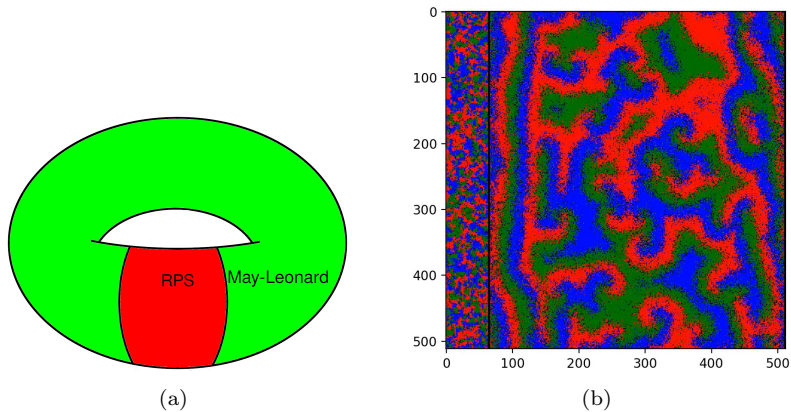

(a)                                      (b)

Fig. 18.  Spatially inhomogeneous rock-paper-scissors model realization with three cyclically competing species on a two-dimensional square lattice with periodic boundary conditions and site exclusion.  (a) Illustration of the torus geometry and division into a smaller patch subject to cyclic Lotka–Volterra rules, and a larger region governed by the May–Leonard reactions.  (b) Monte Carlo simulation snapshot in the (quasi-)stationary state for a $512 \times 512$ square lattice with the rock-paper-scissors patch of width 64 showing its characteristic fluctuating clusters that induce plane waves at both interfaces traveling into the seven times larger May–Leonard region ($\sigma = \lambda = 1$, $D = 0.5$). In the interior bulk of the May–Leonard subsystem one observes its typical spiral structures [116].

In both previous scenarios, stable subsystems, with stationary states that persist in the thermodynamic limit $L \to \infty$, interacted with each other through particle diffusion across the interfaces.  The mutual influences are then limited to within a typical penetration depth set by the correlation length $\xi$, and thus affect only a finite boundary region.  The situation changes drastically when one of the finite patches by itself would be unstable with respect to extinction or fixation.   This finite-size instability with characteristic extinction or fixation time typically scaling exponentially with the particle number in the system $T \sim e^{c\,N}$ can then be suppressed, at least in an extended parameter regime, through a countermanding finite-size effect, namely the continuous supply of individuals from diverse species through the interface of area $\sim L^d$ that traverse the unstable patch in time $t_L \sim L/2v$ with invasion front speed $v$.  If the par-ticle influx is periodically driven by population oscillations at frequency $\omega$ in the stable adjacent regions, it may be roughly estimated as $v \sim \xi \, \omega / 2\pi$. A crude criterion for the induced stabilization of a vulnerable ecosystem would be that the particle import happens fast compared to the extinction time $t_L < T$, i.e., the driving frequency should obey $\omega \, T > \pi \, L / \xi$.[117]

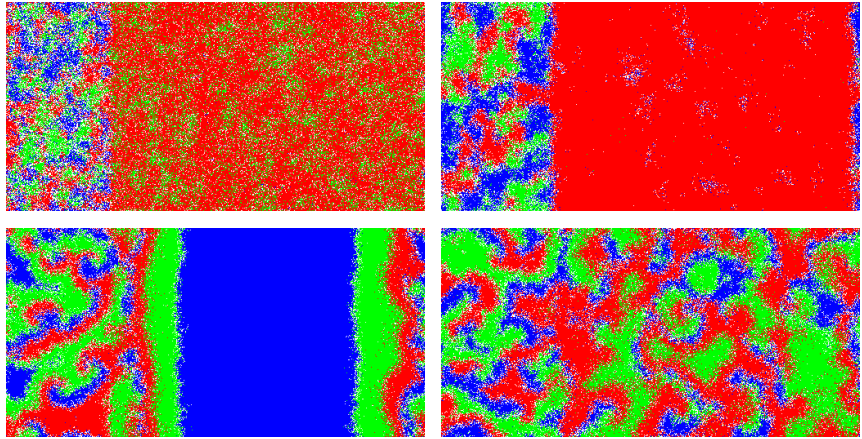

Fig. 19. Monte Carlo simulation snapshots of a two-dimensional inhomogeneous May–Leonard system (periodic boundary conditions) where a (stable) $128 \times 256$ patch with symmetric rates $\sigma = \lambda = 0.2$ and diffusivity $D = 0.8$ is coupled via particle transport across the boundaries to a larger (unstable) $384 \times 256$ region with reduced reaction rate $\lambda' = 0.1$ for the process $A + B \rightarrow A$. At 100 MCS (top left), dominance of one (red) species has developed in the asymmetric patch; at 220 MCS (top right), spiraling clusters have formed in the smaller symmetric region; as time progesses, at 940 MCS (bottom left) planar waves emanate from the stable patch at both interfaces into the asymmetric area, by then fixated at the blue population; ultimately, in the quasi-stationary regime (bottom right), stable spiral structures are continuously seeded by the particle influx into the asymmetric area and persist throughout the simulation's duration. [Figures reproduced with permission from Ref. [96], copyright (2021) by EDP Sciences.]

A striking example is depicted in Fig. 19 which shows two May–Leonard systems adjoined in the same geometric setting as in Fig. 18(a), where the smaller patch on the left-hand side is governed by symmetrically set rates, equal for all three competing species, while the three times larger subsystem on the right is regulated by an asymmetric choice of the predation rates, with $\lambda' = \lambda/2$ for the process $A + B \rightarrow A$ relative to the corresponding reactions for the other two species. The stable symmetric subsystem quickly develops the characteristic spiral structures in the course of the simulation. In contrast, the region with asymmetric predation rates displays pronounced heteroclinic cycles apparent in Fig. 19(b,c) and moreover is sufficiently small to be prone to two-species extinction. Indeed, in Fig. 19(c) its bulk region has effectively fixated at one (blue) population. However, planar invasion fronts are simultaneously excited at both interfaces to the stable May–Leonard subsystem, and quickly invade the entire region, eventually providing a sufficient supply of diverse individuals to stabilize the vulnerable ecology as seen in Fig. 19(d) which features the spiral patterns of the asymmetric system that would also emerge in much larger isolated settings. For this system, one may formulate the stabilization criterion that the induced spiral rotation frequency $\omega$ must exceed the characteristic inverse time scale for structure disruption owing to diffusive particle transport through the spiral arms, which should be proportional to the inverse square of the average correlation length for the three species, $\omega > D/\langle\xi^2\rangle$.[96] Similar observations can be made in spatially inhomogeneous May–Leonard system combinations where the asymmetry is prescribed in the diffusivity rather than reactivity, as well as in diffusively coupled two-species Lotka–Volterra predator models where one patch is rendered finite-size unstable through increasing its predation rate or carrying capacity.[117]

## 7. Conclusions and outlook

The stochastic spatially extended Lotka–Volterra model for predator-prey competition and coexistence provides a paradigmatic example for the potentially severe quantitative as well as qualitative shortcomings of the (crudest) mean-field description in terms of deterministic coupled nonlinear rate equations in many nonequilibrium systems. Internal fluctuations as well as external variability in the model parameters are seen to have profound influence on the dynamics of the system. Perhaps most strikingly, intrinsic randomness in conjunction with spatio-temporal correlations induced by the binary reaction kinetics induce spontaneous pattern formation in dimensions $d \leq 4$. Hence, exploring fluctuation and correlation effects beyond the mean-field description constitutes an important task in physical, chemical, biological, ecological, and sociological models, as they may cause major changes to the dynamics and induce novel emergent phenomena. Even when the rate equations provide a qualitatively adequate picture, there usually will be quantitative parameter renormalizations in stochastic nonlinear dynamical systems.

A continuum field theory representation of the stochastic master equation represents a powerful analytical tool to properly incorporate fluctu-ation and correlation effects, but of course, explicit calculations can be performed only for comparatively simple models. Individual-based Monte Carlo simulations are an extremely useful tool to understand the temporal evolution, possible emergence of structures, and quantitative characterization of interacting stochastic many-particle systems, and may provide helpful guidance for further mathematical analysis. However, a full scan through the entire parameter space is rarely feasible, and aside from potential algorithmic artifacts, numerical data are naturally subject to statistical errors and finite-size limitations that cannot always be overcome within reasonable time frames and cost boundaries. The optimal strategy to enhance our understanding of complex systems in nature will therefore always be to utilize a combination of all state-of-the-art methods at our disposal.

Our exploration of spatial predator-prey models over the past two decades turned out vastly more intriguing than originally anticipated, revealing many unexpected phenomena and pertinent features connected to quite distinct physical or biological systems. The summary in this chapter inevitably focuses on our group's contributions and likely but inadvertently does injustice to many other researchers' important work, for which the author sincerely apologizes. This research topic covers a broad interdisciplinary area, which has been exciting to engage in, but which has also posed its specific challenges owing to the dispersal of the relevant literature and the distinct terminology and fundamental approaches in different scientific fields. There still remains much to be accomplished in refining our methods and conceptual grasp, and extending them to more complicated scenarios and especially observational and experimental realizations. I already look forward to seeing and learning about these novel future developments.

## Acknowledgements

The author deeply appreciates the organizers' kind invitation to participate in the Ising Lectures series in May 2023. This book chapter is dedicated to the Ukrainian people and their valiant struggle for freedom in peace.

The author gratefully acknowledges his insightful and productive collaborators in the reported and related work: Lazarus Arnau, Matthew Asker, Timo Aspelmeier (deceased), Nicholas Butzin, John Cardy, Jacob Carroll, Sheng Chen, Lauren Childs, Udaya Sree Datla, Oliviér Deloubrière, Shengfeng Deng, Kenneth Distefano, Ulrich Dobramysl, Kim Forsten-Williams, Erwin Frey, Ivan Georgiev, Yadin Goldschmidt, Manoj Gopalakrishnan, Qian He, Bassel Heiba, Lluís Hernandez Navarro, Henk Hilhorst,

Haye Hinrichsen, Martin Howard, Hans-Karl Janssen, Julianna Kelley, Wei Li, Weigang Liu, Jerôme Magnin, William Mather, Mauro Mobilia, Ruslan Mukhamadiarov, Riya Nandi, Michel Pleimling, Priyanka, Matthew Raum, Beth Reid, Alastair Rucklidge, Beate Schmittmann, Gunter Schütz, Franz Schwabl (deceased), Shannon Serrao, Sara Shabani, Mohamed Swailem, Reda Tiani, Steffen Trimper, Fréderic van Wijland, Ben Vollmayr-Lee, Mark Washenberger, Louie Hong Yao, Canon Zeidan, and Royce Zia.

This research was in part supported by the U.S. National Science Foundation, Division of Materials Research under Award No. DMR-0308548; the U.S. National Science Foundation, Division of Mathematical Sciences under current Award No. NSF DMS-2128587; the U.S. Department of Energy, Office of Basic Energy Sciences, Division of Materials Sciences and Engineering under Award No. DEFG02-09ER46613; and was sponsored by the U.S. Army Research Office and was accomplished under Grant No. W911NF17-1-0156. The views and conclusions contained in this document are those of the author and should not be interpreted as representing the official policies, either expressed or implied, of the Army Research Office or the U.S. Government.

# Index

*Index*